\documentclass[11pt]{article}

\usepackage{graphicx}
\usepackage{epsfig}
\usepackage{cite}
\usepackage{amssymb}
\usepackage{amsmath}
\usepackage{dsfont}
\usepackage{multirow}
\usepackage{ulem}
\usepackage{soul}
\usepackage{sidecap}
\include{preamble}

\usepackage{url}
\usepackage{hyperref}
\usepackage{amsfonts}

\newcommand{\be}{\begin{equation}}

\newcommand{\ee}{\end{equation}}
\newcommand{\bea}{\begin{eqnarray}}
\newcommand{\eea}{\end{eqnarray}}
\newcommand{\bi}{\begin{itemize}}
\newcommand{\ei}{\end{itemize}}
\newcommand{\ben}{\begin{enumerate}}
\newcommand{\een}{\end{enumerate}}

\newcommand{\lp}{\left(}
\newcommand{\rp}{\right)}

\def\frac#1#2{{{#1}\over {#2}}}
\def\gsim{\mathrel{\rlap{\lower4pt\hbox{\hskip1pt$\sim$}}
    \raise1pt\hbox{$>$}}}         
\def\lsim{\mathrel{\rlap{\lower4pt\hbox{\hskip1pt$\sim$}}
    \raise1pt\hbox{$<$}}}         

\newcommand{\tot}{\mathrm{tot}}

\newcommand{\draft}[1]{}

\def\MS{\hbox{$\overline{\rm MS}$}}

\def\QMS{Q$_0$\MS}

\usepackage{amssymb}

\begin{document}




\begin{flushright}
IFUM-962-FT\\
\end{flushright}
\begin{center}
{\Large \bf
HERA data and DGLAP evolution:\\ theory and phenomenology}
\vspace{0.8cm}

Fabrizio~Caola, Stefano~Forte and Juan~Rojo, 

\vspace{1.cm}
{\it  Dipartimento di Fisica, Universit\`a di Milano and
INFN, Sezione di Milano,\\ Via Celoria 16, I-20133 Milano, Italy \\}

\bigskip
\bigskip

{\bf \large Abstract:} 
\end{center}
We examine critically the evidence for deviations from next-to-leading
order perturbative DGLAP evolution in HERA data. We briefly review the
status of perturbative small-$x$ resummation and of global
determinations of parton distributions. We then show that the
geometric scaling properties of HERA data are consistent with DGLAP
evolution, which is also strongly supported by the double asymptotic
scaling properties of the data. We finally show that backwards
evolution of parton distributions into the low $x$, low $Q^2$ region
shows evidence of deviations between the observed behaviour and the
next-to-leading order predictions. These deviations cannot be
explained by missing next-to-next-to-leading order perturbative terms,
but are consistent with perturbative small-$x$ resummation.












\section{DGLAP evolution in the LHC era}
\label{sec:intro}
Perturbative QCD is a quantitatively tested theory which describes in
a very accurate way a vast body of data, and it is at the basis of
physics at colliders such as the LHC~\cite{Altarelli:2008zz}. The
DGLAP equations, i.e. the
renormalization-group equations which govern the scale dependence of
parton distributions are, together with asymptotic freedom and
factorization, a cornerstone of the theory, both in terms of
phenomenological success and theoretical foundation. Specifically,
they are the tool which allows us to combine information on nucleon
structure from a variety of data, and use it for predictions at the
LHC. 

The current frontier in perturbative QCD is systematically going from
the second to the third perturbative order, namely from
next-to-leading (NLO) to next-to-next-to-leading order (NNLO). In view
of this, it is of utmost importance for both theory and phenomenology
to understand whether in any given kinematic region an order-by-order
perturbative approach is sufficient.
There is now considerable theoretical evidence that at sufficiently
high center-of-mass energies this approach may break down, and thus as
this region is approached one should resum to all orders
perturbative corrections which are
logarithmically enhanced in the ratio $x$ of the hard scale to
center-of-mass energy
--- the so--called small-$x$ resummation.  However, conclusive experimental
evidence for such resummation effects in the data is lacking, partly
due to the fact that the relevant effects are difficult to disentangle
from model and theoretical assumptions.

It is the purpose of this contribution to review,  update and put in
context a recent attempt to provide some such evidence. 
The outline of this contribution is as follows. In
 Sect.~\ref{sec:smallxres}
we briefly review the status of linear small-$x$
resummation. Then in Sect.~\ref{sec:fixedorder} we present
the state of the art of fixed-order DGLAP and global PDF
analysis and its implications for the LHC. Finally in
 Sect.~\ref{sec:deviations} we  
present a technique designed to identify deviations from
NLO
DGLAP evolution in the data, and apply it to recent global parton fit.
We will specifically see that previous evidence of deviations from NLO
DGLAP is considerably strengthened by the recent
precise
combined HERA-I determination of deep-inelastic structure functions.

\section{Small-$x$ resummation}
\label{sec:smallxres}

As well known, deep--inelastic partonic cross sections and parton
splitting functions
receive large corrections in the small-$x$ limit due to the presence of powers 
of $\alpha_s\log x$ to all orders in the perturbative 
expansion~\cite{Catani:1990eg,Catani:1994sq}.
This suggests dramatic effects from yet higher
orders, so the success of NLO perturbation theory 
at HERA has been
for a long time very hard to explain.
In the last several years this situation 
has been clarified~\cite{Altarelli:2003hk,Altarelli:2005ni,Altarelli:2008aj,Ciafaloni:2003rd,Ciafaloni:2003kd,Ciafaloni:2007gf},
showing that, once the full resummation procedure
accounts for running coupling effects, gluon exchange symmetry and
other physical constraints, the effect of
the resummation of terms which are enhanced at small-$x$ is
perceptible but moderate --- comparable in size to typical NNLO fixed
order corrections in the HERA region. 

\begin{figure}[ht]
\centering
\includegraphics[width=0.48\textwidth]{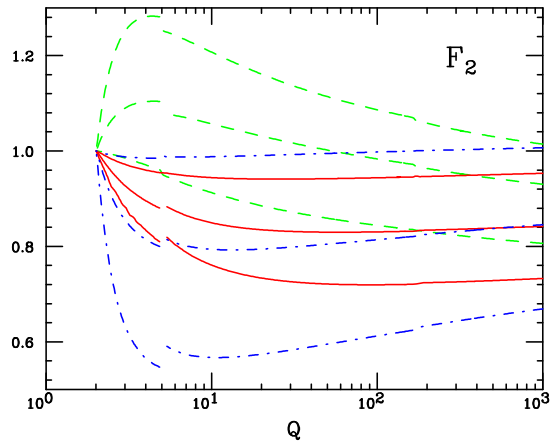}
\caption{\small Ratio of the resummed and NNLO prediction
to the NLO
fixed order
 for the singlet $F_2$ deep--inelastic structure function.
 The curves are:   
fixed order perturbation theory NNLO (green, dashed);
resummed NLO  in \QMS\ scheme (red, solid),
resummed NLO in the \MS\ scheme (blue, dot-dashed).
In each case, the three curves shown correspond to fixed $x=10^{-2}$,
$x=10^{-4}$, $x=10^{-6}$, with the smallest $x$ value showing the
largest deviations. 
}\label{plot_f2resum}
\end{figure}
For phenomenology, it is necessary to resum not only evolution
equations, but also hard partonic coefficient functions. The relevant
all--order coefficients have been computed for
DIS~\cite{Catani:1990eg,Catani:1994sq}, and more recently  
for several LHC processes such as 
heavy quark production~\cite{Ball:2001pq}, 
Higgs production~\cite{Marzani:2008az,Marzani:2008ih},
Drell-Yan~\cite{Marzani:2008uh,Marzani:2009hu} 
and prompt photon production~\cite{Diana:2009xv,Diana:2010ef}. 
These coefficients can be used for a full resummation of physical
observables, by suitably combining them with resummed DGLAP evolution
and accounting for running--coupling effects, so as to maintain perturbative
independence of physical observables of the choice of factorization
scheme~\cite{Ball:2007ra}.

This program has been carried out for the first time 
for deep--inelastic scattering in 
Ref.~\cite{Altarelli:2008aj}, and more recently for prompt photon
production~\cite{Rojo:2010gv}, which makes resummed phenomenology for these
processes possible. In particular in Ref.~\cite{Altarelli:2008aj}
results have been presented for the ratio of resummed to unresummed
NLO
deep--inelastic structure functions. These ratios are shown in
Fig.~\ref{plot_f2resum} for two choices of the resummed factorization
scheme discussed in Ref.~\cite{Altarelli:2008aj},
and compared to analogous ratio of the fixed
order NNLO to NLO. They were determined under the hypothesis that
the structure functions $F_2$ and $F_L$ are kept fixed for all $x$ at
$Q_0=2$~GeV: this  models the situation in which parton distributions
are determined  at the scale $Q_0$, and one then sees the change in
prediction when going from NLO to NNLO, or from NLO to unresummed.
In Sect.~\ref{sec:deviations} we will present evidence
for departures from the NLO  which appear to be consistent with this figure.

\section{Fixed order DGLAP: from HERA to LHC}
\label{sec:fixedorder}

\begin{figure}[ht]
\begin{center}
\epsfig{width=0.65\textwidth,figure=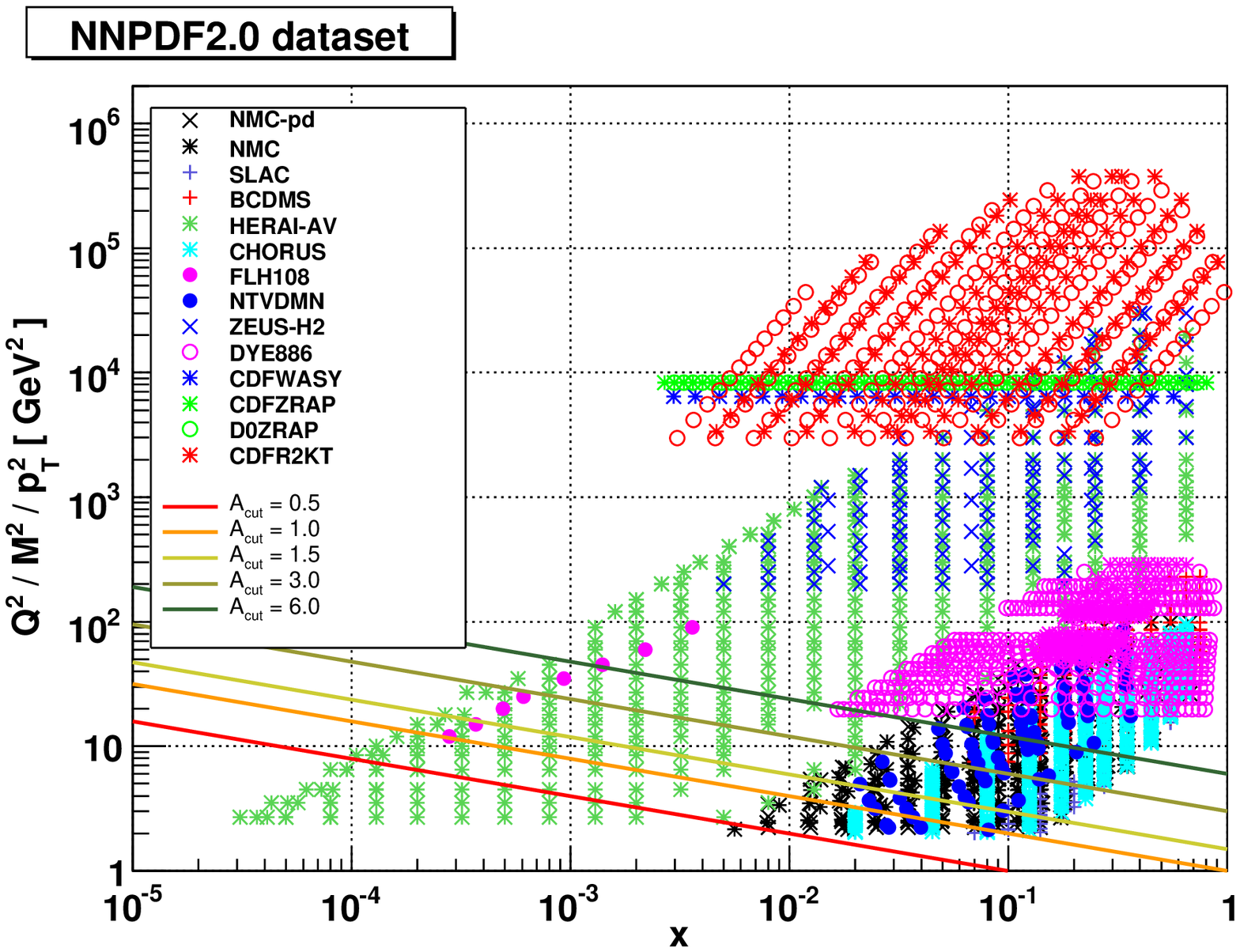}
\epsfig{width=0.34\textwidth,figure=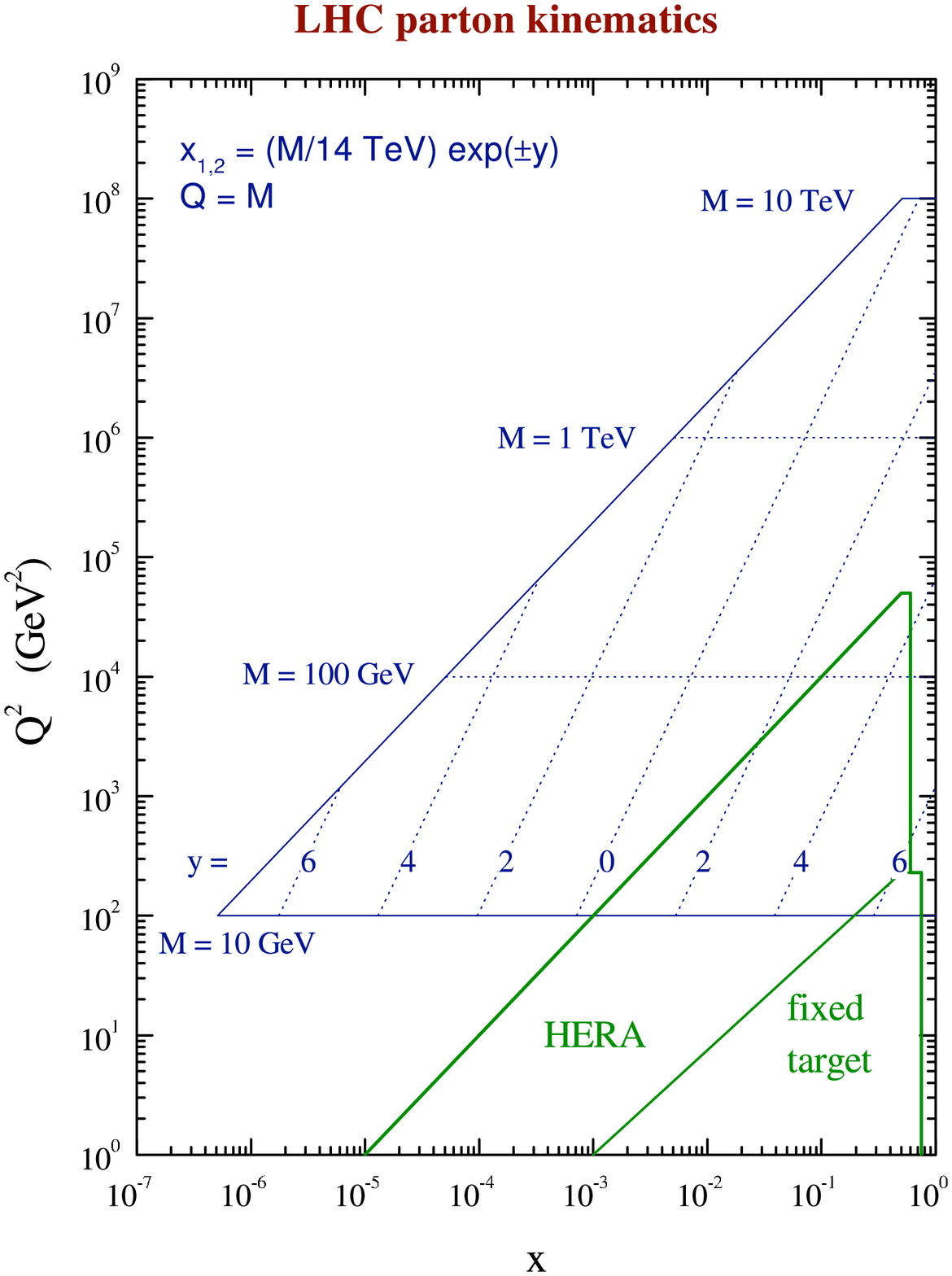}
\caption{ \small Left: Experimental data
used in  the NNPDF2.0 global analysis;  the series of $A_{\rm cut}$  
kinematic cuts is discussed  in Sect.~\ref{sec:deviations}. Right: 
LHC kinematical region.
\label{fig:dataplot}} 
\end{center}
\end{figure}

Fixed--order DGLAP evolution is an integral ingredient of any PDF
determination. 
Currently, the
most comprehensive PDF sets are obtained from a global analysis of
hard-scattering data from a variety of processes like deep--inelastic
scattering, Drell-Yan and weak vector boson production and 
collider jet production.
In such global analysis, QCD factorization and DGLAP evolution are
used to relate experimental data to a common set of PDFs.
Three groups produce such global analysis and provide regular updates
of these: NNPDF~\cite{Ball:2010de}, CTEQ~\cite{Nadolsky:2008zw, Lai:2010nw} 
and 
MSTW~\cite{Martin:2009iq}. The typical dataset included in one of 
such global analysis is shown in Fig.~\ref{fig:dataplot}. We also
show in Fig.~\ref{fig:dataplot}
the kinematic range which is available at the LHC
as compared to that covered by present experimental data: 
extrapolation to larger $Q^2$ from the current data region is possible 
thanks to DGLAP evolution.

A significant  advance in global PDF
analysis in the recent years has been the development
of the NNPDF
methodology~\cite{f2ns,f2p,DelDebbio:2007ee,Ball:2008by,Rojo:2008ke,Ball:2009mk,Ball:2009qv,Ball:2010de}. NNPDF provides a 
determination  of PDFs and their uncertainty which is  independent of the
choice of data set, and which has been shown in benchmark
studies~\cite{Dittmar:2009ii} to behave in a statistically consistent
way when data are added or removed to the fit. Also, because of the use of a
Monte Carlo approach, the NNPDF methodology is easily amenable 
to the use of standard
statistical analysis tools. The most updated NNPDF analysis is
 NNPDF2.0~\cite{Ball:2010de}, a
global fit to all relevant DIS and hadronic hard scattering data.

\begin{figure}[ht]
\begin{center}
\epsfig{width=0.48\textwidth,figure=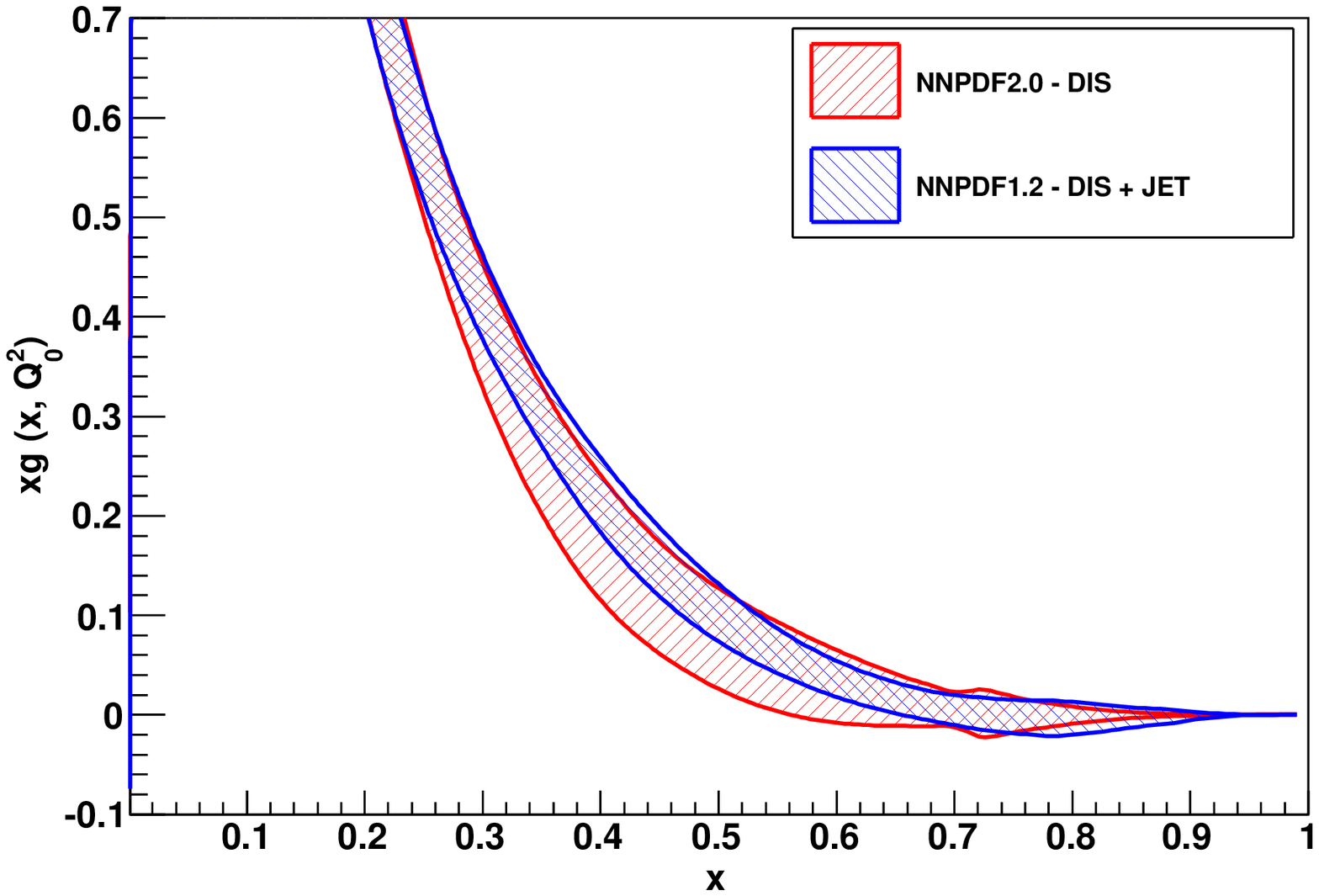}
\epsfig{width=0.48\textwidth,figure=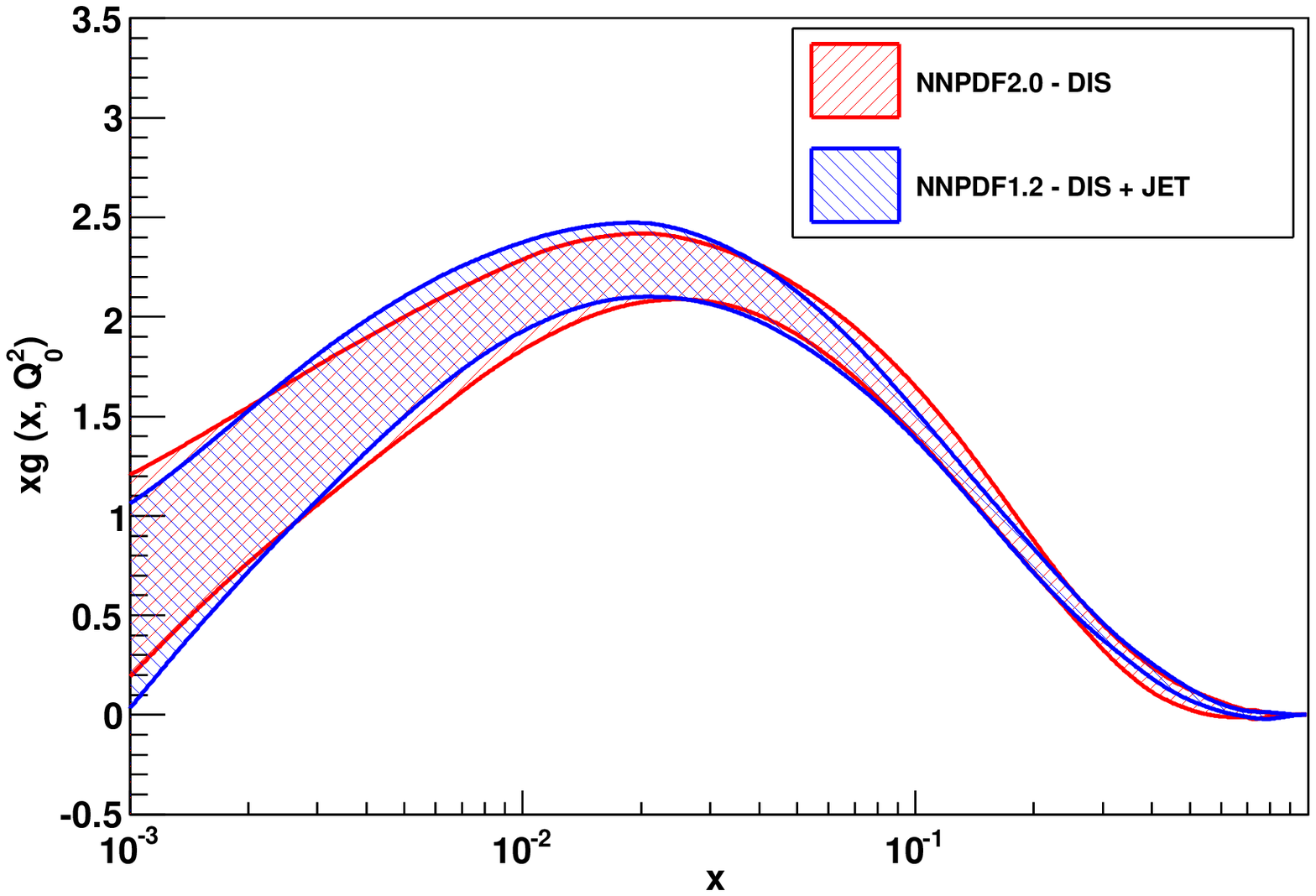}
\caption{ \small 
Impact on the gluon distribution of the inclusion of jet data  in a fit
  with DIS data.
\label{fig:gluon}} 
\end{center}
\end{figure}

Comparing the effect of individual datasets on a global fit such as 
NNPDF2.0 allows detailed studies of QCD factorization,
DGLAP evolution, and 
the compatibility between DIS and hadronic data.
A very stringent test is obtained by comparing the results of
a fit to DIS data only to that of DIS+jet data
(Table~\ref{tab:estdataset1}). Indeed, it turns out that jet data,
which are at much higher scale, are well predicted by PDFs determined
from lower scale DIS data. Furthermore, the gluon extracted from the
DIS--only fit, which is essentially determined from DGLAP scaling
violations, turns out to agree very well with that determined when jet
data are also included (see Fig.~\ref{fig:gluon}): upon inclusion of
the jet data, the uncertainty decreases without a significant change
in central value.

\begin{table}
\centering
\scriptsize
\begin{tabular}{|c|c|c|c|}
\hline  
Fit & 2.0 DIS &  2.0 DIS+JET & NNPDF2.0 \\
\hline 
\hline
$\chi^{2}_{\tot}$ &  1.20 & 1.18  &  1.21  \\
\hline
\hline
 NMC-pd    & 0.85&  0.86 & 0.99 \\
\hline
NMC             &  1.69 &  1.66  & 1.69 \\
\hline
SLAC            & 1.37 & 1.31& 1.34 \\
\hline
BCDMS           & 1.26 &  1.27 & 1.27 \\
\hline
HERAI        & 1.13 & 1.13  & 1.14 \\
\hline
CHORUS          & 1.13& 1.11& 1.18 \\
\hline
NTVDMN          &  0.71&  0.75 & 0.67 \\
\hline
ZEUS-H2         & 1.50 &  1.49 & 1.51 \\
\hline
DYE605          & {\it 7.32} & {\it 10.35}  & 0.88 \\
\hline
DYE866          &  {\it 2.24}&  {\it 2.59}  & 1.28 \\
\hline
CDFWASY         & {\it 13.06}& {\it 14.13} &  1.85 \\
\hline
CDFZRAP         & {\it 3.12}&  {\it 3.31} & 2.02  \\
\hline
D0ZRAP          &  {\it 0.65}&  {\it 0.68} & 0.47 \\
\hline
CDFR2KT         & {\it 0.91}&  0.79 & 0.80 \\
\hline
D0R2CON         &  {\it 1.00} & 0.93 & 0.93 \\
\hline
\end{tabular}
\caption{\small  The $\chi^2$ for
individual experiments included in NNPDF2.0 fits with
DIS data only, DIS and jet data only,
and the full DIS, jet and Drell-Yan data set.
For each fit, values of the
$\chi^2$ for data not included in the fit are shown in italic.  The
value of $\chi^2_{\rm tot}$ in the first line 
does not include these data.}\label{tab:estdataset1} 
\end{table}

\begin{figure}
\begin{center}
\epsfig{width=0.48\textwidth,figure=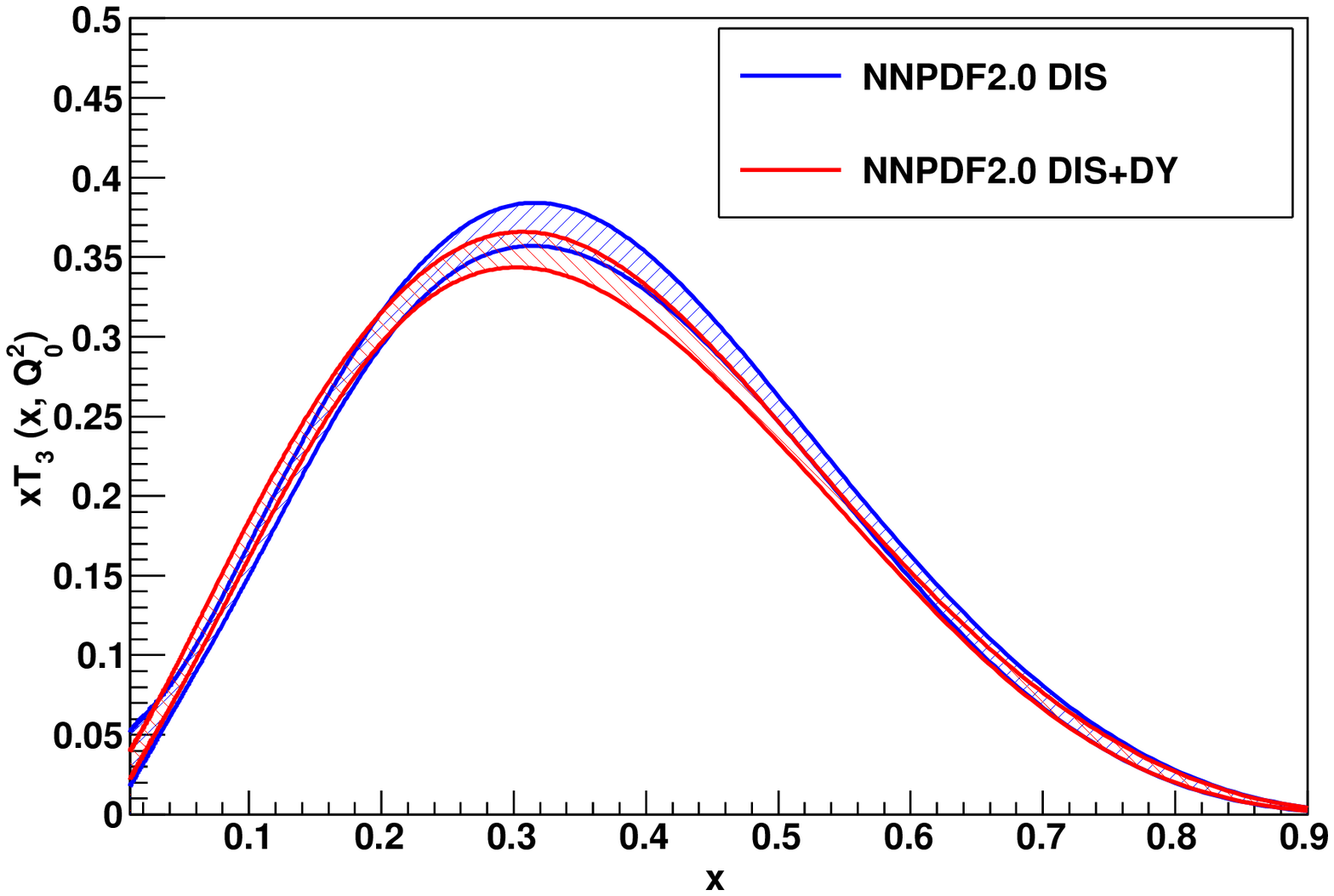}
\epsfig{width=0.48\textwidth,figure=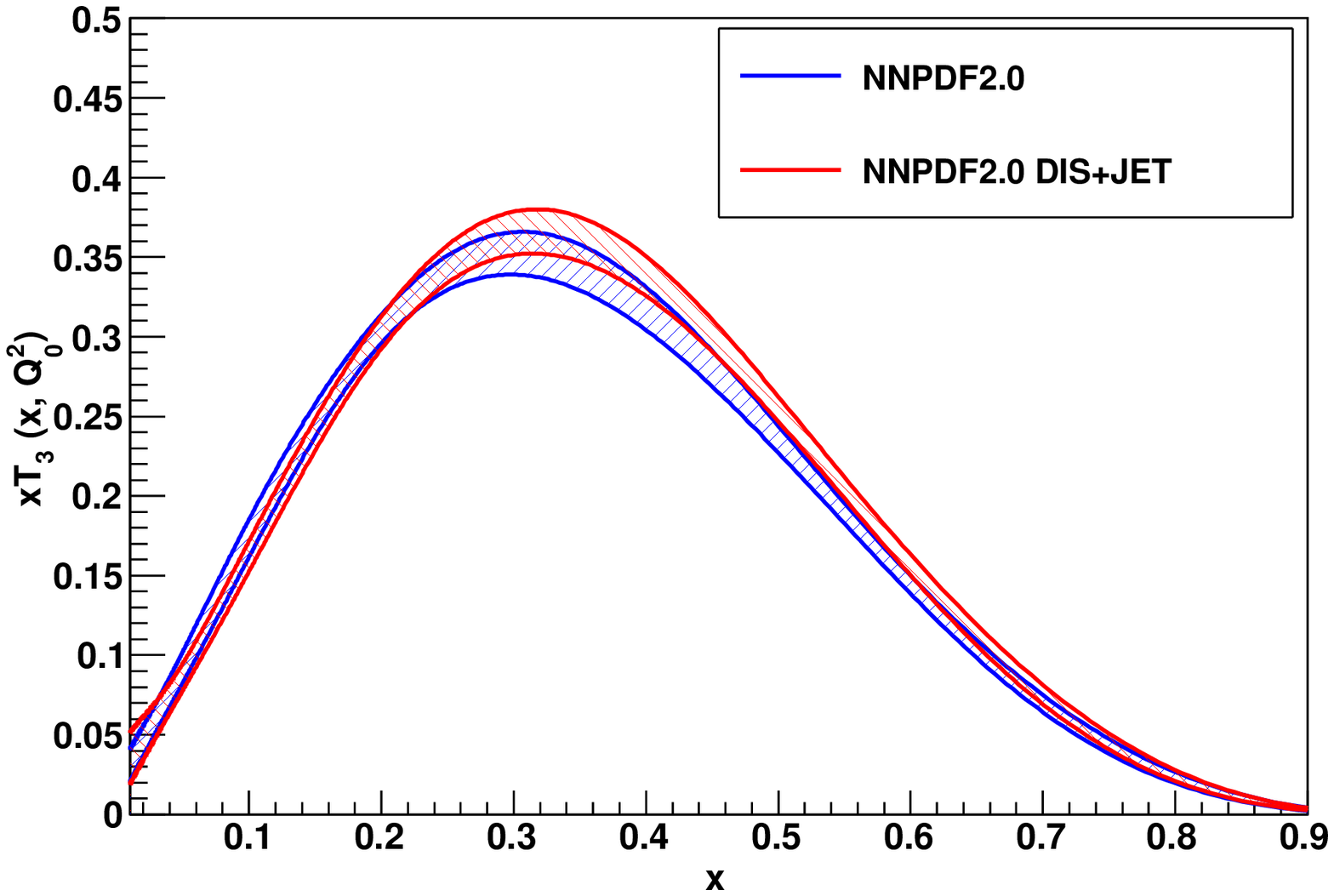}
\epsfig{width=0.48\textwidth,figure=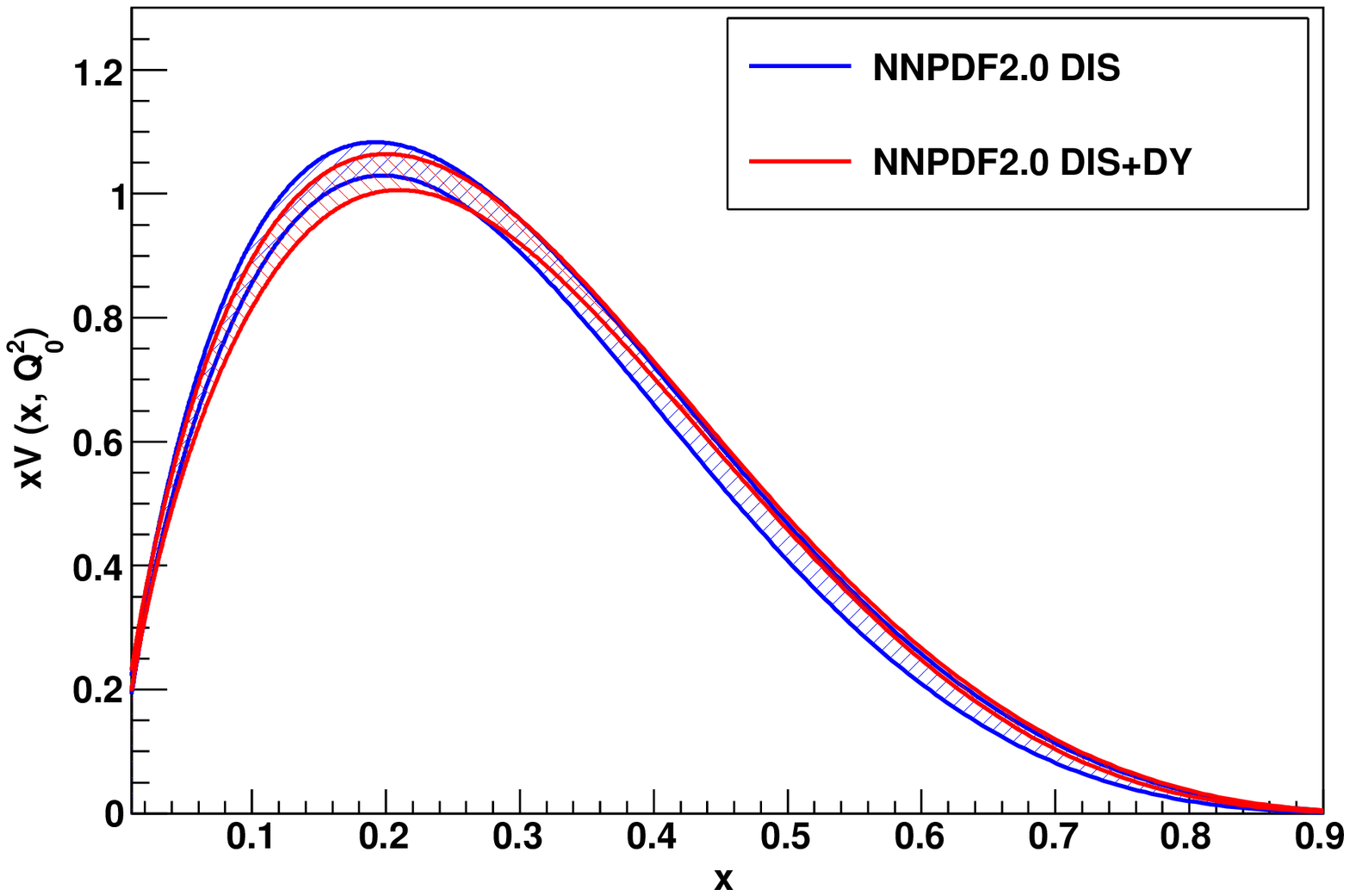}
\epsfig{width=0.48\textwidth,figure=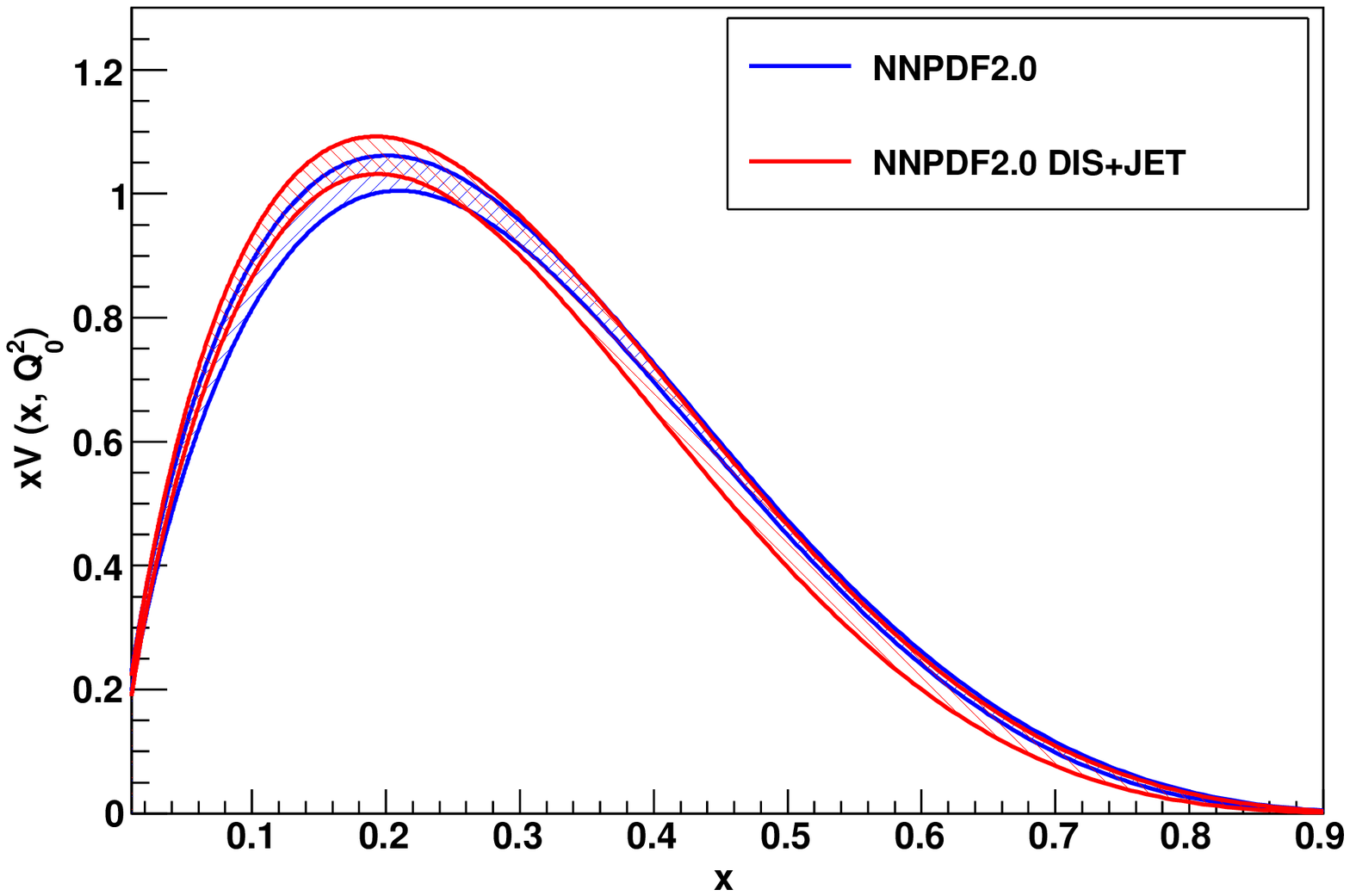}
\epsfig{width=0.48\textwidth,figure=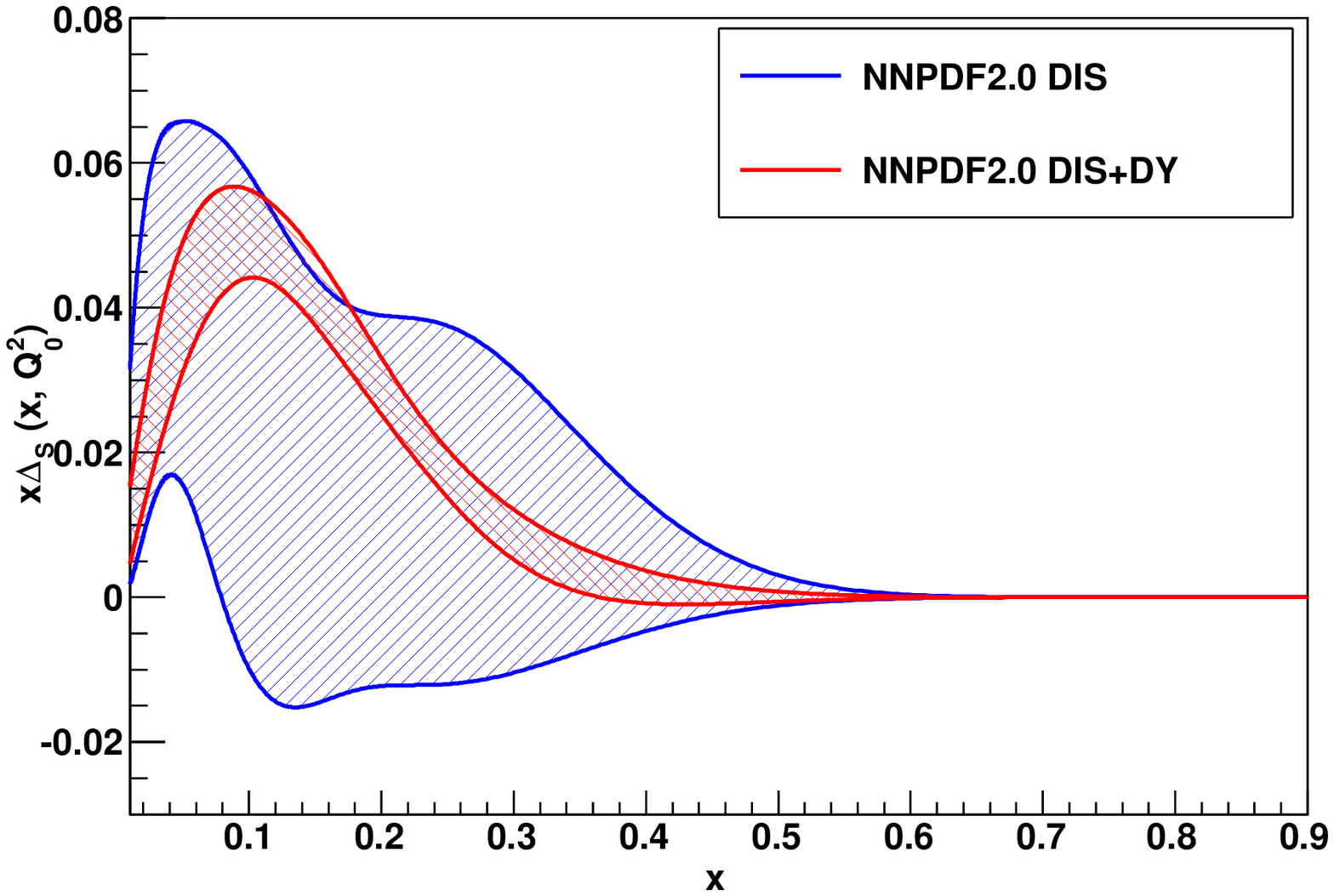}
\epsfig{width=0.48\textwidth,figure=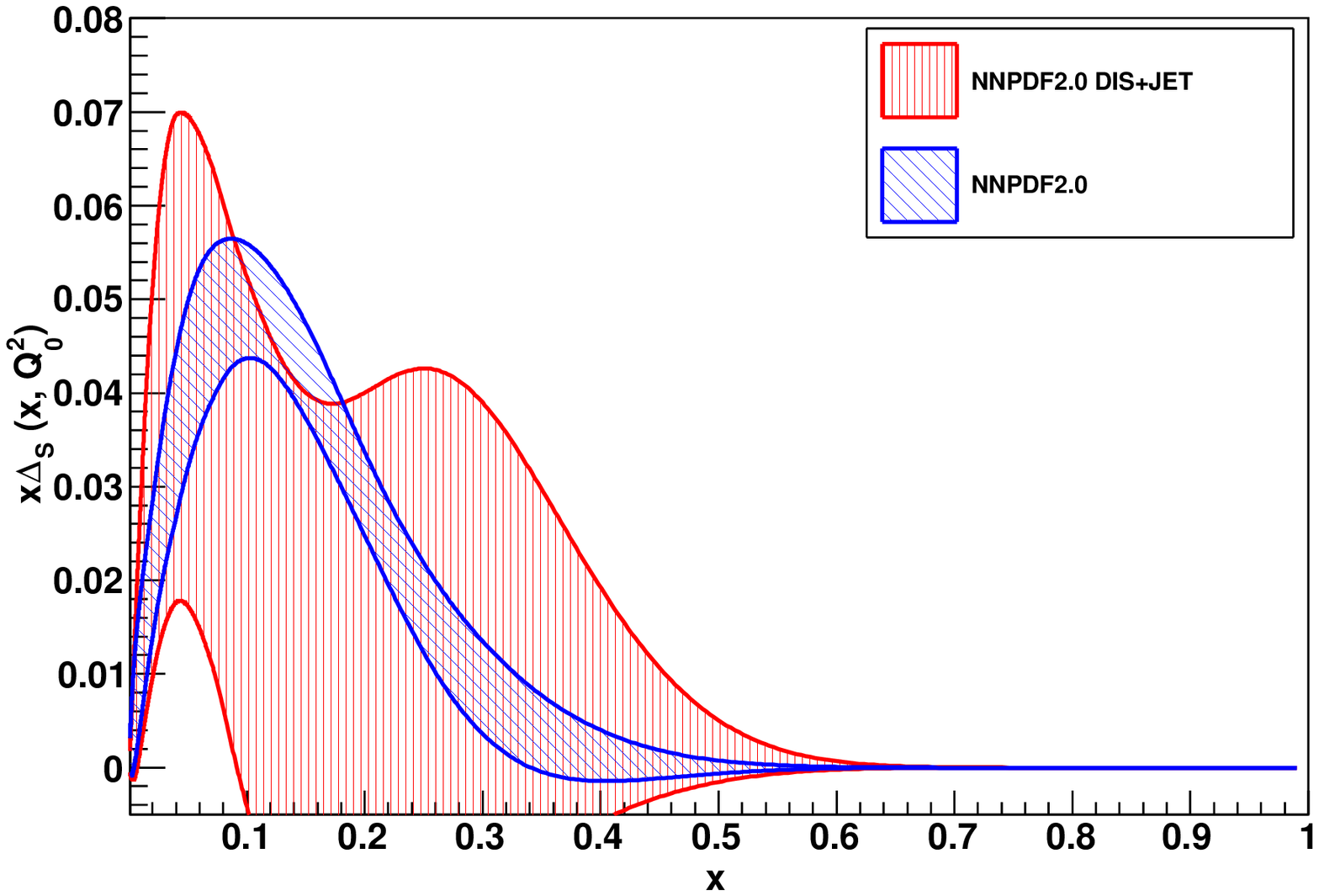}
\epsfig{width=0.48\textwidth,figure=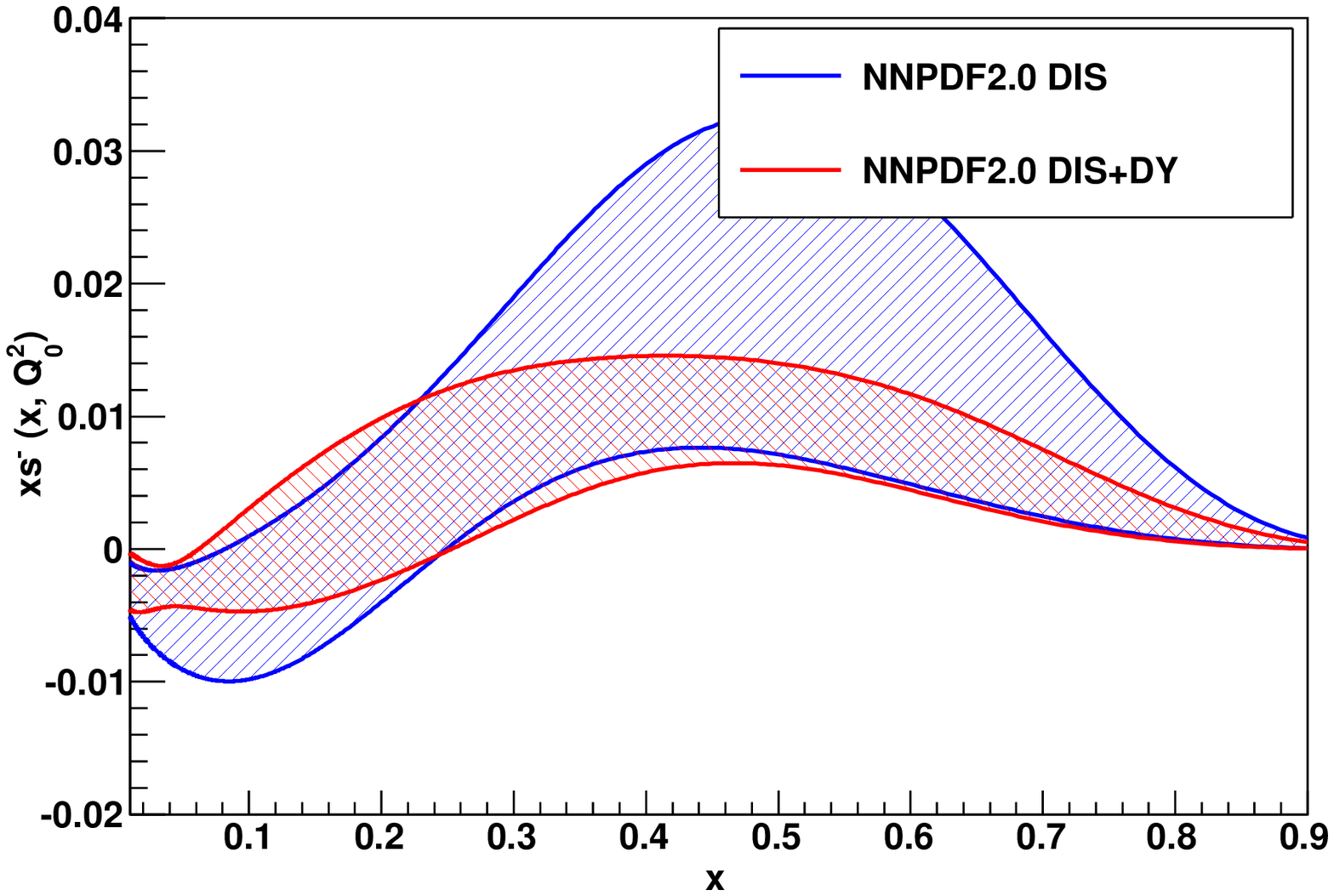}
\epsfig{width=0.48\textwidth,figure=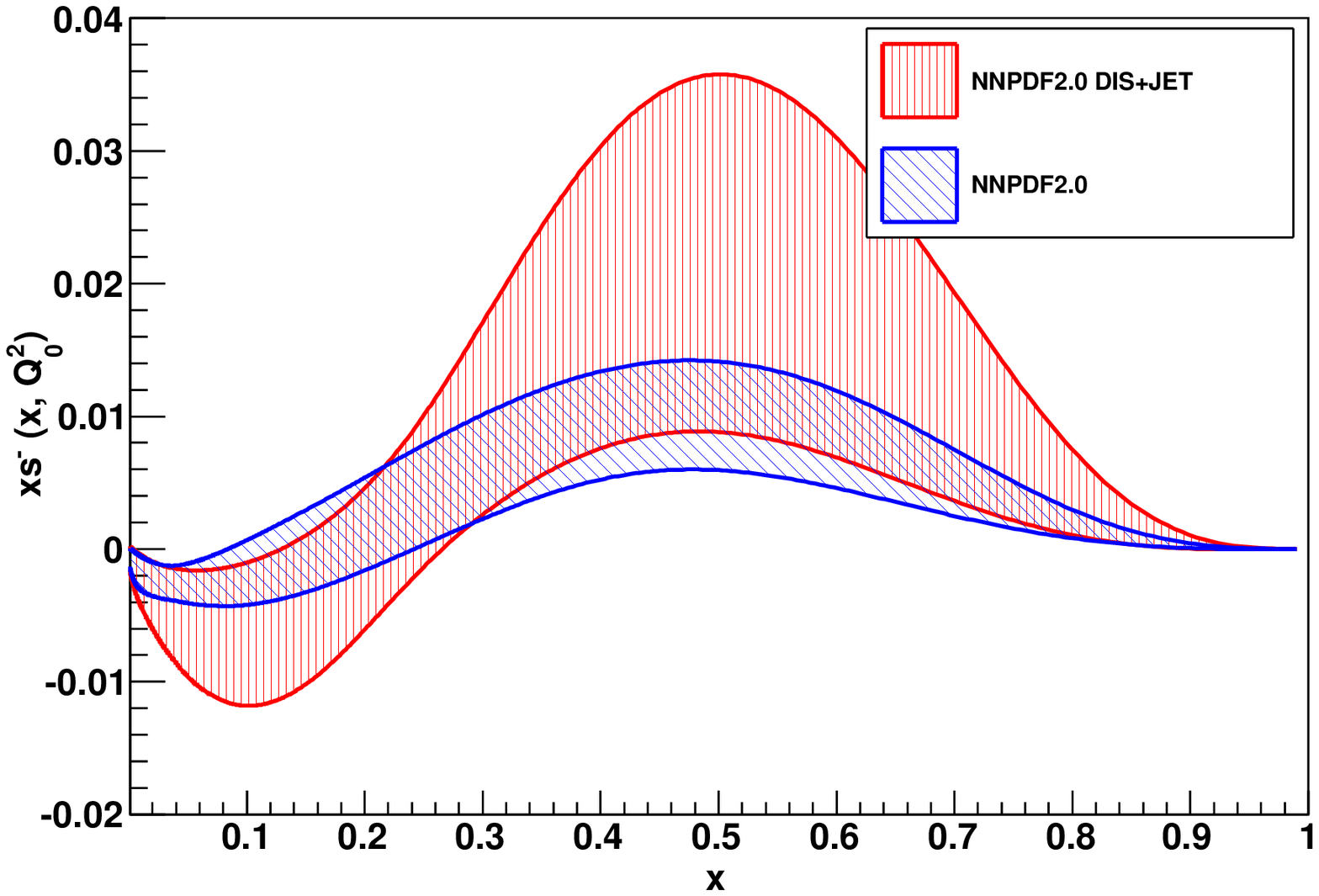}
\caption{ \small Impact of the inclusion of Drell--Yan data in a fit
  with DIS data only (left), and in a fit with DIS and jet data
  (right). From top to bottom, isotriplet, total valence, $\bar d-\bar
  u$ and $s-\bar s$ PDFs
  are shown in each case.
\label{fig:commute}} 
\end{center}
\end{figure}

Further consistency checks are obtained by comparing the effect of the
inclusion of a specific dataset (such as Drell-Yan) to different
datasets (such as DIS, or DIS+jets). If there was any inconsistency
between different sets,
the impact of the new data would be different according to whether
they are added to data they are or are not consistent with. No such
differences are observed (see Fig.~\ref{fig:commute}). Because the
various sets are at different scales and related through DGLAP evolution
this also provides a strong check of its accuracy.

\section{Deviations from DGLAP in HERA data}
\label{sec:deviations}

We have seen that 
NLO DGLAP is extremely successful in describing in
a consistent way all relevant hard scattering data. 
On the other hand, there are several theoretical 
indications that at small $x$ and/or at small $Q^2$ the
NLO DGLAP might undergo sizable corrections due to
leading--twist small-$x$ perturbative resummation, or non linear 
evolution,  parton saturation and other higher twist effects.
Even if it is unclear in which kinematical regime
these effects 
should become relevant, it  is likely that eventually
they should become relevant, specifically in order to prevent
violations of unitarity.

\begin{figure}
\centering
\epsfig{file=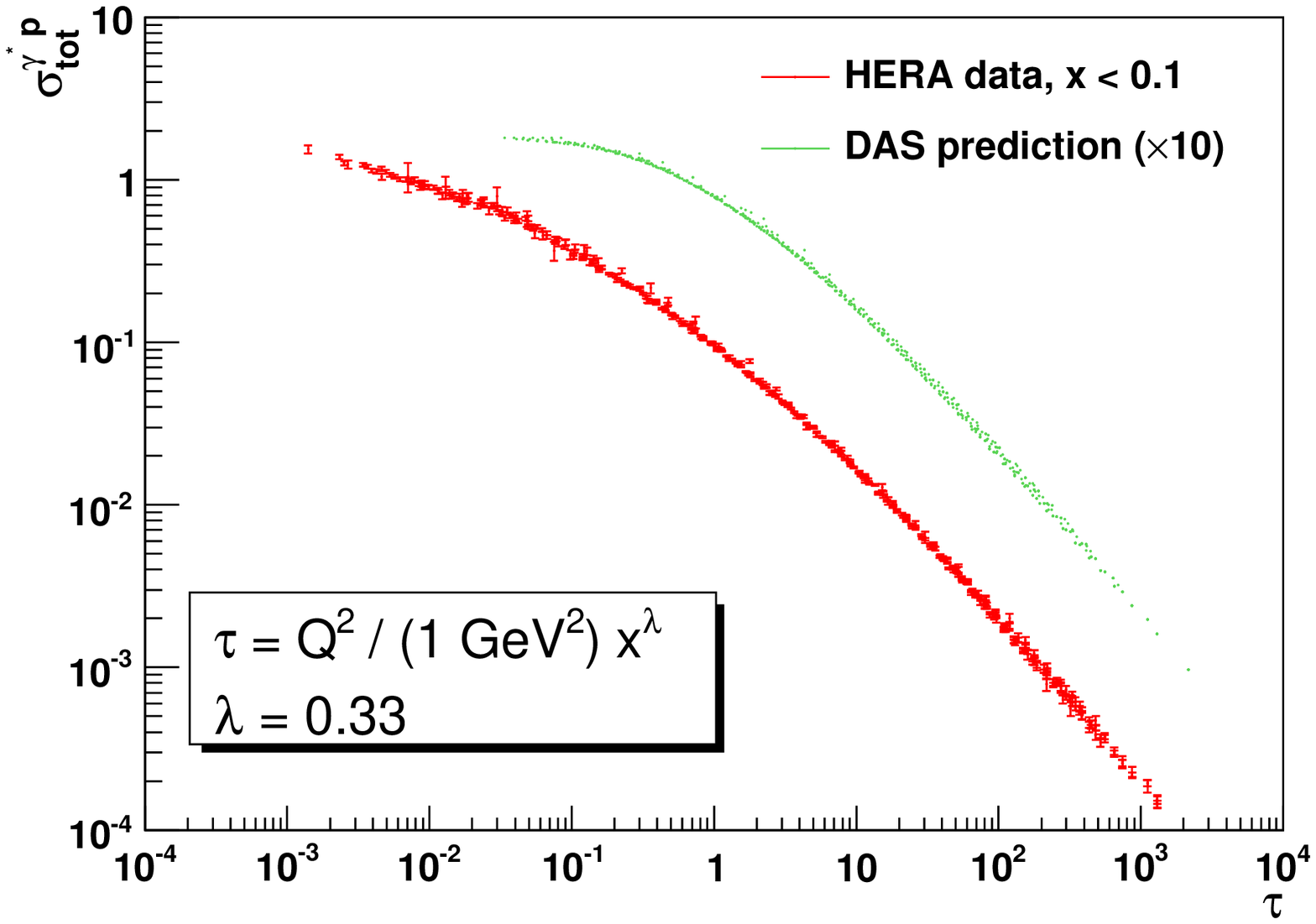, scale = 0.35}
\epsfig{file=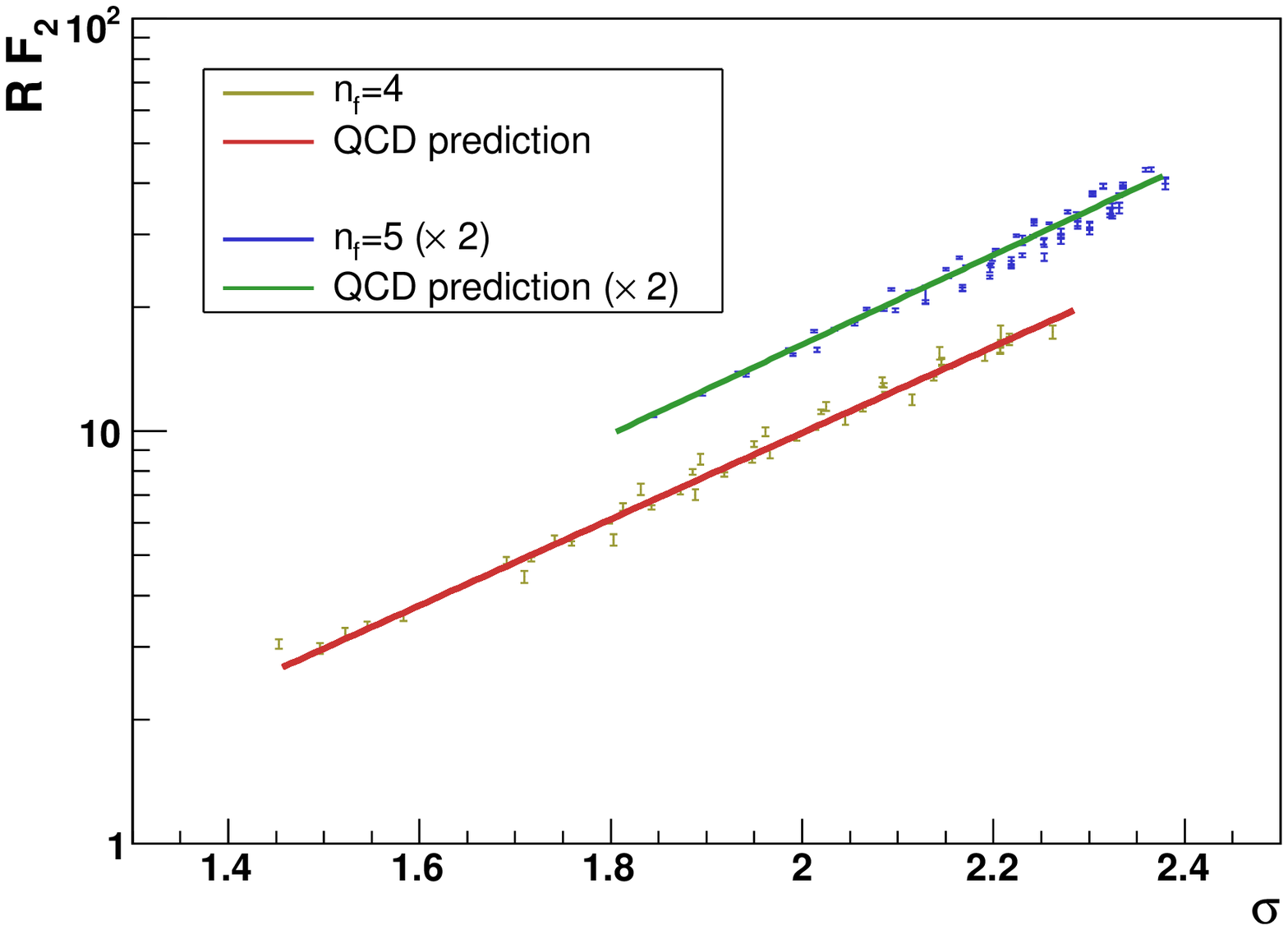, scale=0.35}
\caption{\small Left: geometric scaling of the 
HERA-I data and of the (Double Asymptotic Scaling)
small--$x$ solution
of the LO DGLAP evolution equation for a
 a flat boundary condition at $Q_0^2=1$~GeV$^2$. 
Only points with $Q^2>10$~GeV$^2$ are included in the DAS curve, which
is offset for clarity. 
Right: the rescaled proton $F_2$ structure function plotted
as a function of the DAS variables $\sigma\equiv \sqrt{\ln 1/x \ln \left(
\ln t/t_0\right)}$, with $t=\ln Q^2/\Lambda_{QCD}^2$ 
and $Q_0=1$~GeV. The points correspond to the same data and  the
curves 
corresponds to the same prediction as in the right plot. The two
curves include data
with $10$~GeV$^2\ge Q^2\ge 25$~GeV$^2$ and $x<0.01$ ($n_f=4$) 
or $Q^2\ge 25$~GeV$^2$ and $x<0.07$ ($n_f = 5$).}
\label{das_plot}
\label{gs_plot}
\end{figure}
When trying to trace these effects, one should beware of the
possibility that putative signals of deviation might in fact be
explained using standard NLO theory.
An example of this situation
is the so--called  geometric scaling~\cite{Stasto:2000er} , which 
is often thought to
provide unequivocal evidence for saturation. This is the prediction,
common to many saturation models,
that
 DIS cross sections, 
at small-$x$ depend only on the 
single variable 
\be\tau(x,Q^2) = \lp Q^2/Q_0^2 \rp\cdot (x/x_0)^\lambda, \label{eq:tau}
\ee
rather than on $x$ and $Q^2$ separately. However, it turns out that
geometric scaling Eq.~(\ref{eq:tau})
is also generated by linear DGLAP evolution ~\cite{Caola:2008xr}: 
fixed order DGLAP evolution evolves any (reasonable) 
boundary condition 
into a geometric scaling form. Furthermore, the 
scaling exponent $\lambda$ Eq.~(\ref{eq:tau}) obtained in such way
 agrees very well 
with the experimental value.

Following Ref.~\cite{Caola:2008xr},
in Fig.~\ref{gs_plot} we compare the scaling behaviour of the
HERA data 
and the LO 
small-$x$ DGLAP evolution of a flat boundary condition: the DGLAP 
prediction scales even better than data. Note that
the  accurate combined HERA-I
dataset~\cite{H1ZEUS:2009wt} are used here, and that   
the scaling behaviour persists also at larger $x\lsim 0.1$, where it is 
unlikely that it is related to saturation.
Clearly, this shows that geometric scaling is not sufficient to
conclude that fixed--order DGLAP fails. However, one may wonder
whether small-$x$ LO DGLAP is phenomenologically relevant.

To answer this, in Fig.~\ref{gs_plot} we also show a comparison of the
data to the so--called double asymptotic scaling
(DAS)~\cite{Ball:1994du} form, obtained from the small-$x$ limit of
the LO DGLAP solution. The agreement between data and theory is so
good that one can see the change in DAS slope when the number of
active flavours goes from $n_f=4$ to $n_f=5$: so, while on the one hand
geometric scaling cannot discriminate  between pure DGLAP and
saturation, double asymptotic scaling provide evidence that the data
follow the predicted DGLAP behavior in most of the HERA region.

However,  deviations from DGLAP evolution can be investigated
exploiting
the more discriminating and sensitive 
framework of global PDF fits. The key idea in this kind of
analysis is to perform global fits only in the large-$x$, large-$Q^2$ region, 
where NLO DGLAP is certainly reliable. This way one can determine ``safe'' 
parton distributions which are not contaminated by possible non-DGLAP effects. 
These ``safe'' PDFs are then evolved backwards into the potentially
``unsafe'' low-$x$ and low-$Q^2$ kinematic 
region, and
used to compute physical observables, which are compared with data. 
A deviation between the predicted and observed behaviour in this
region can then provide
a signal for effects beyond NLO DGLAP. 
Since possible deviations are small, this kind of studies is
meaningful only on statistical grounds, hence a reliable estimate of PDFs
uncertainties and theoretical biases is mandatory. The NNPDF framework
provides useful tools for this kind of investigations.

In~\cite{Caola:2009iy} such an analysis was performed using  the 
NNPDF1.2 PDF set~\cite{Ball:2009mk}, and it did provide some evidence
for deviations from NLO DGLAP. Here, we update this analysis using
the 
NNPDF2.0 PDF set discussed in Sect.~\ref{sec:fixedorder}.
In comparison to NNPDF1.2, NNPDF2.0
also includes hadronic data (fixed target Drell-Yan production, 
collider weak boson production  and 
collider inclusive jets), and the
 combined HERA-I dataset~\cite{H1ZEUS:2009wt} replaces previously less
 accurate data from ZEUS and H1.
The inclusion of the very accurate
combined HERA-I dataset has the potential to increase the
significance of the observed deviations from DGLAP, while
the presence of hadronic data allows
stringent tests of the global compatibility of the
NLO DGLAP framework, as discussed in Sect.~\ref{sec:fixedorder}. A
further
difference between NNPDF1.2 and NNPDF2.0 is an improved treatment
of normalization uncertainties based on the so--called  $t_0$ 
method~\cite{Ball:2009qv},
which avoids the biases of other commonly used methods to deal
with normalization uncertainties.

The ``safe'' region,
where non--DGLAP  effects are likely to be negligible, is defined as
\begin{equation}
\label{qs}
Q^2 \ge A_{\rm cut} \cdot x^\lambda,
\end{equation}
with $\lambda=0.3$. This definition has the feature of only
considering unsafe small-$x$ data if their scale is low enough, with
the relevant scale raised as $x$ is lowered; its detailed shape is
inspired by saturation and resummation studies.
We have performed fits with only data which pass the cut
Eq.~(\ref{qs}) included, with a variety of choices for  $A_{\rm cut}$,
shown in  Fig.~\ref{fig:dataplot}. Results depend smoothly on
$A_{\rm cut}$.

\begin{figure}
\begin{center}
\includegraphics[width=0.48\textwidth]{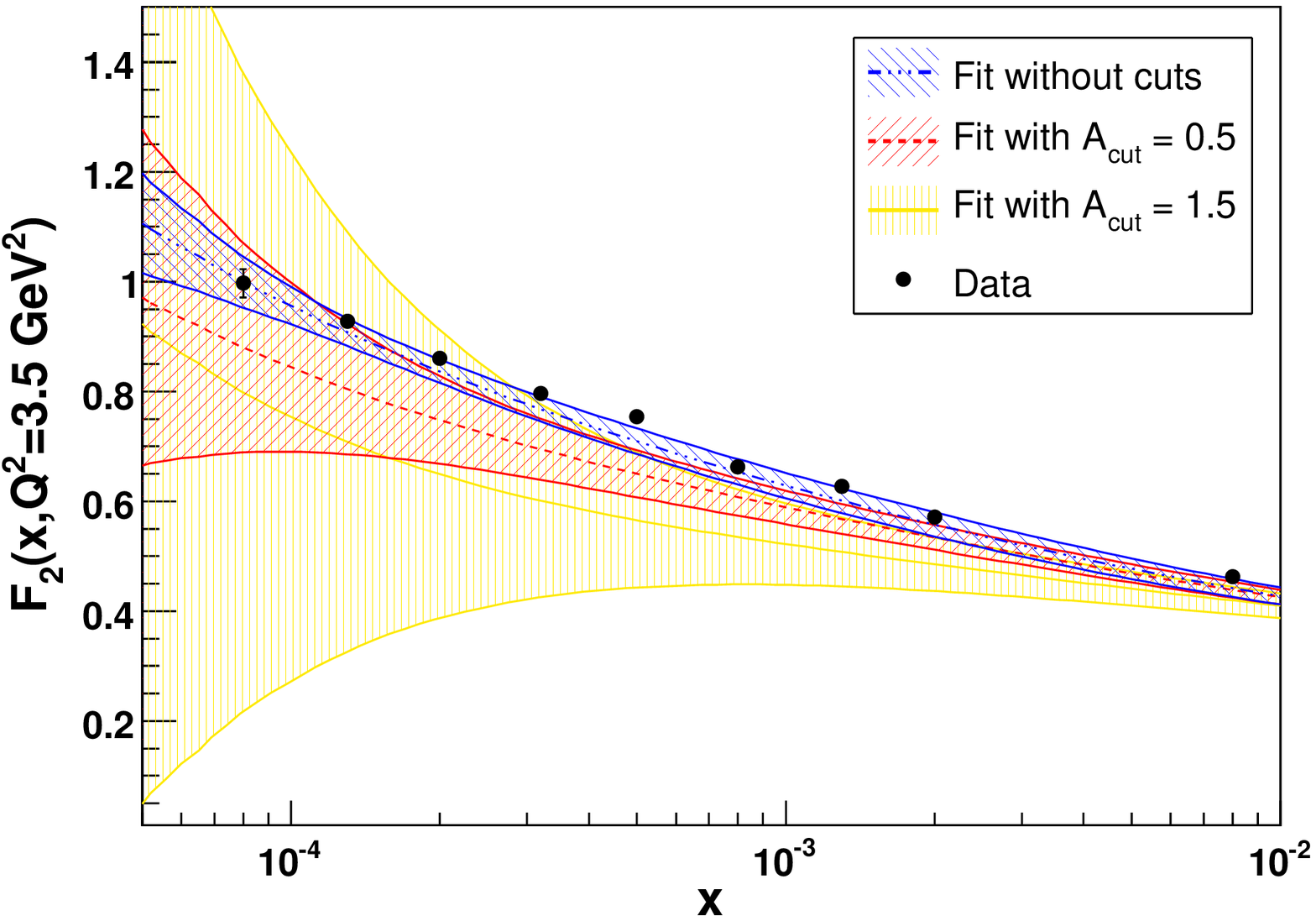}
\includegraphics[width=0.48\textwidth]{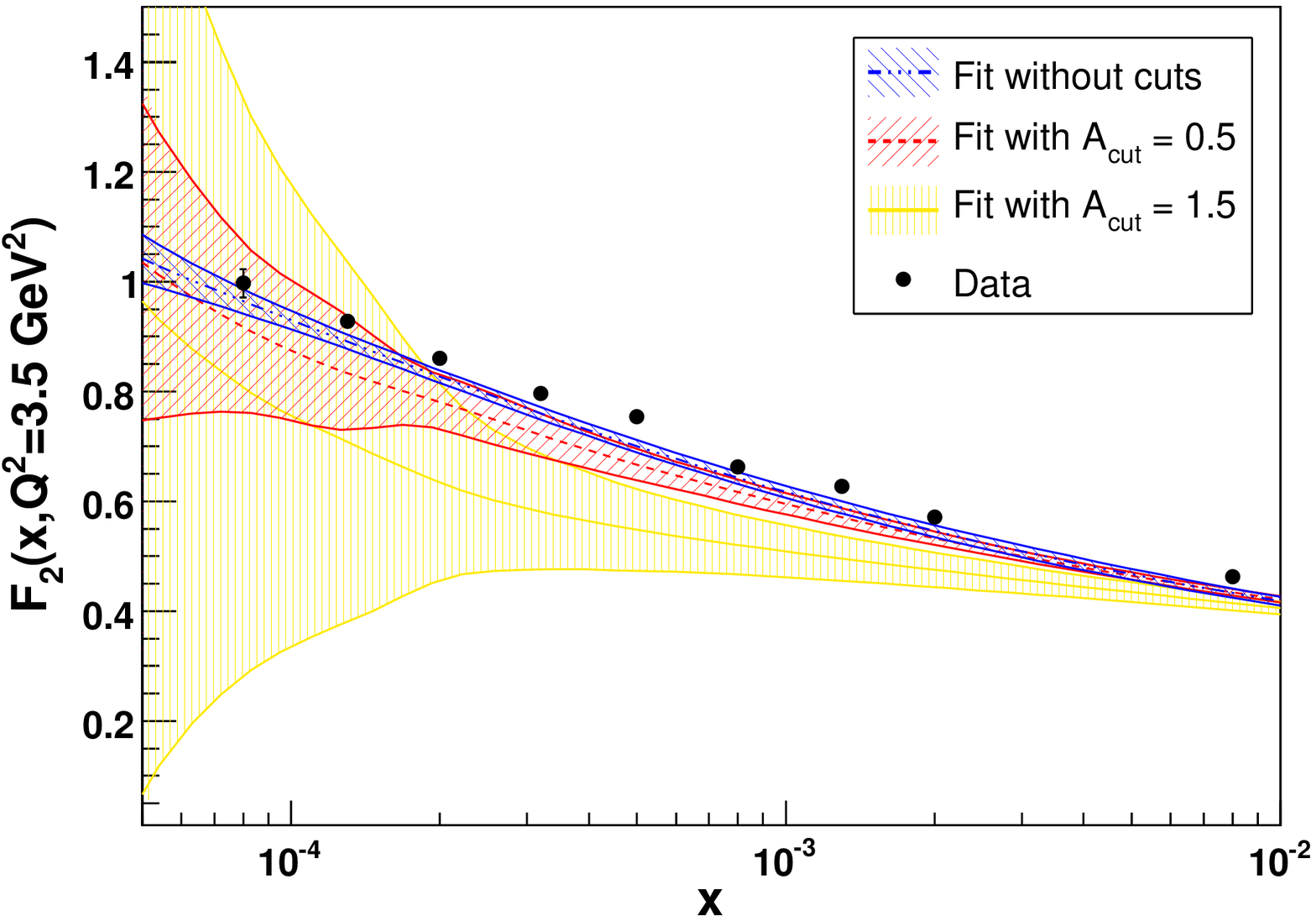}
\end{center}
\caption{\small 
The proton structure function $F_2(x,Q^2)$ 
computed using NNPDF1.2 (left) or NNPDF2.0 (right) 
PDFs obtained from fits with different 
values of $A_{\rm cut}$.}\label{f2_15_pic}\label{f2_35_pic}
\end{figure}

As a first  test,
 we have computed the proton structure function 
$F_2$ and compared it with data (see Fig.~\ref{f2_35_pic}) 
at  $Q^2=3.5~$GeV$^2$, where a
 significant $x$ range falls below the cut (compare with Fig.~\ref{fig:dataplot}).
Clearly,
the prediction obtained from backward evolution of the data above the cut
exhibits a systematic downward trend.  This deviation, which becomes more and
more apparent as $A_{\rm cut}$  is raised, is visible but marginal
when  the NNPDF1.2 set based on old HERA data is used, but it becomes
rather more significant when using NNPDF2.0 and new HERA
data. Interestingly, with old HERA data the uncut fit agrees well with
the data, showing that whatever  the possible deviation between data
and theory, it is absorbed by the PDFs. This is no longer possible
when the more precise combined HERA data are used: in such case, even
when no cut is applied, the theory cannot reproduce the data fully.
This suggests that low-$x$ and  
$Q^2$ NLO DGLAP evolution is stronger than the scale dependence seen
in the data.

In order to quantify this observed deviation from NLO DGLAP, we introduce the 
statistical distance 
\begin{equation}\label{stat_dist}
d_{\rm stat}(x,Q^2) \equiv \frac{F_{\rm data}-F_{\rm fit}}{\sqrt{\sigma^2_{\rm data}+\sigma^2_{\rm fit}}},
\end{equation}
i.e. the difference between the observable $F_{\rm data}$ and
the NLO DGLAP prediction $F_{\rm fit}$ in unit of their combined
uncertainties  
$\sigma_{\rm fit}$, $\sigma_{\rm data}$. 
Note that  $d_{\rm stat}\sim 1$ corresponds to a
1--sigma effect.

\begin{figure}
\centering
\epsfig{file=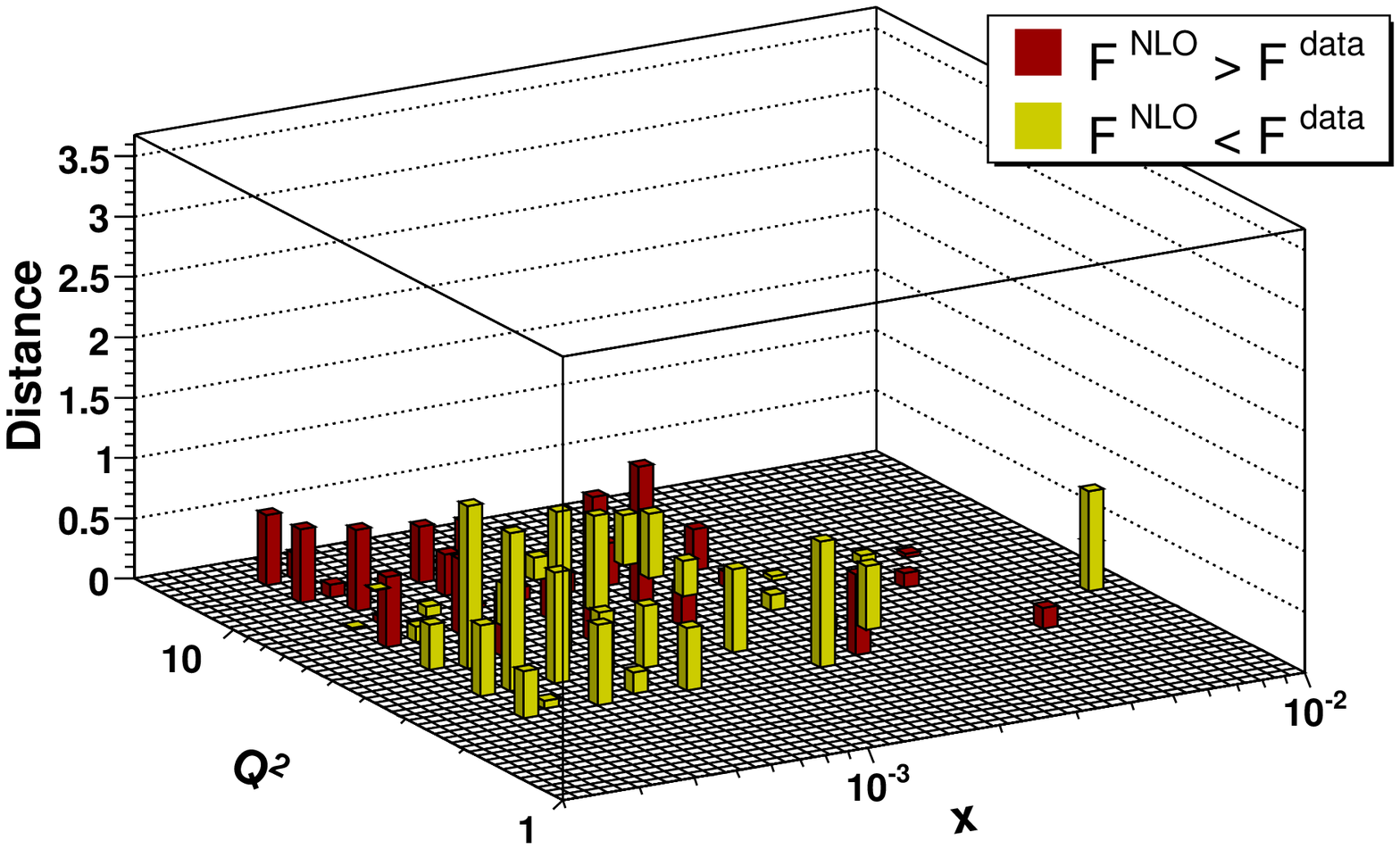, scale=0.32}
\epsfig{file=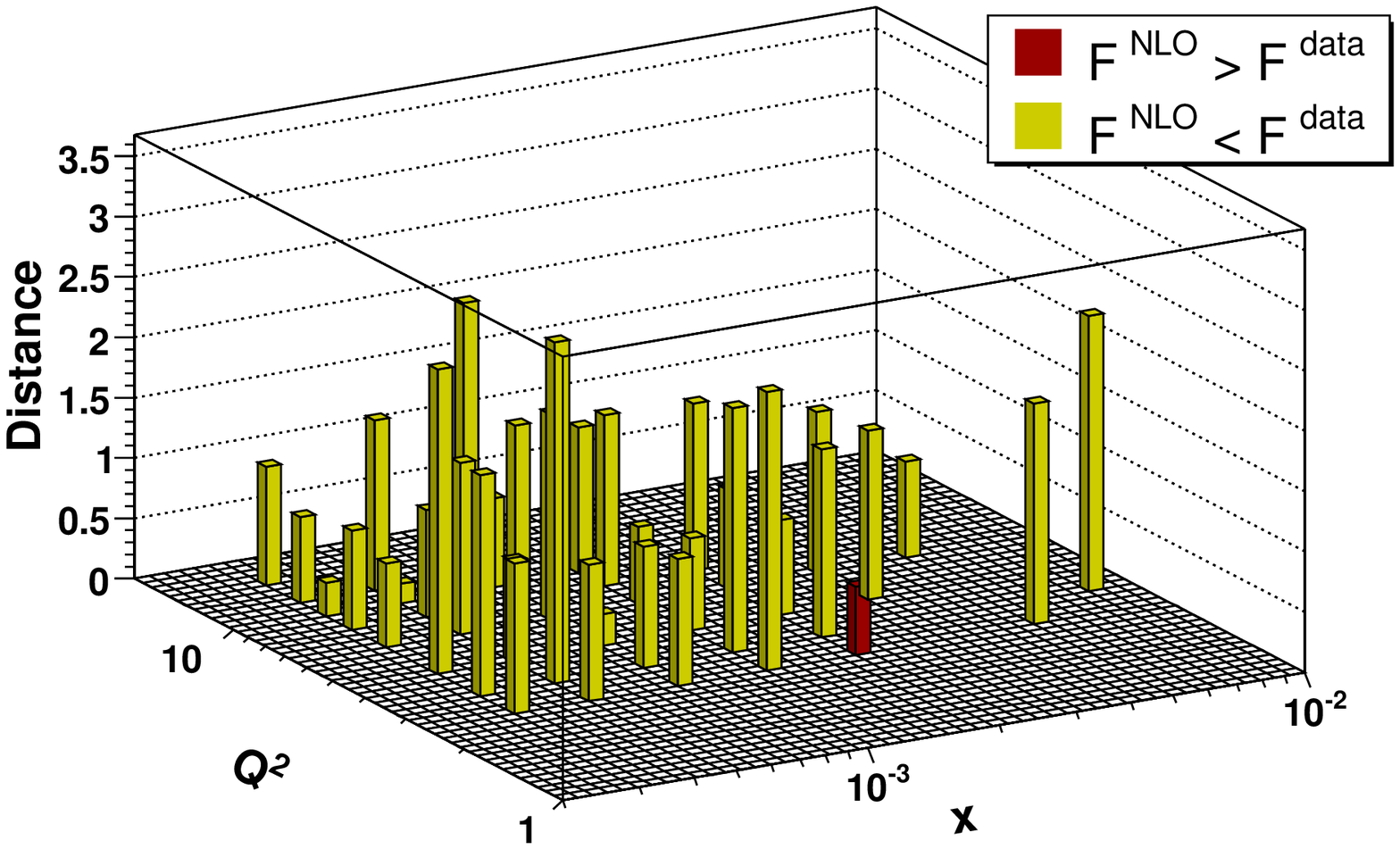, scale=0.32}
\epsfig{file=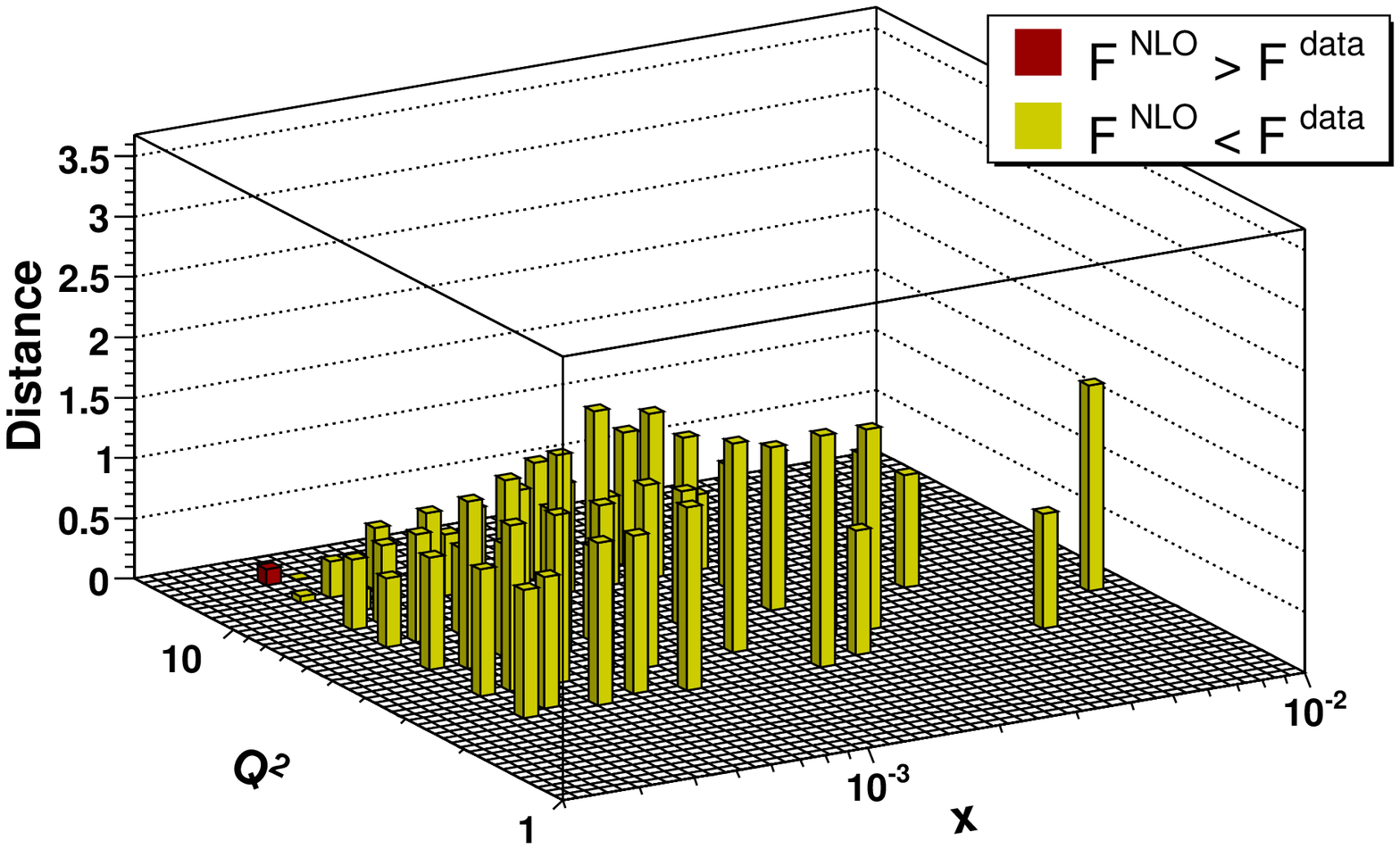, scale=0.32}
\epsfig{file=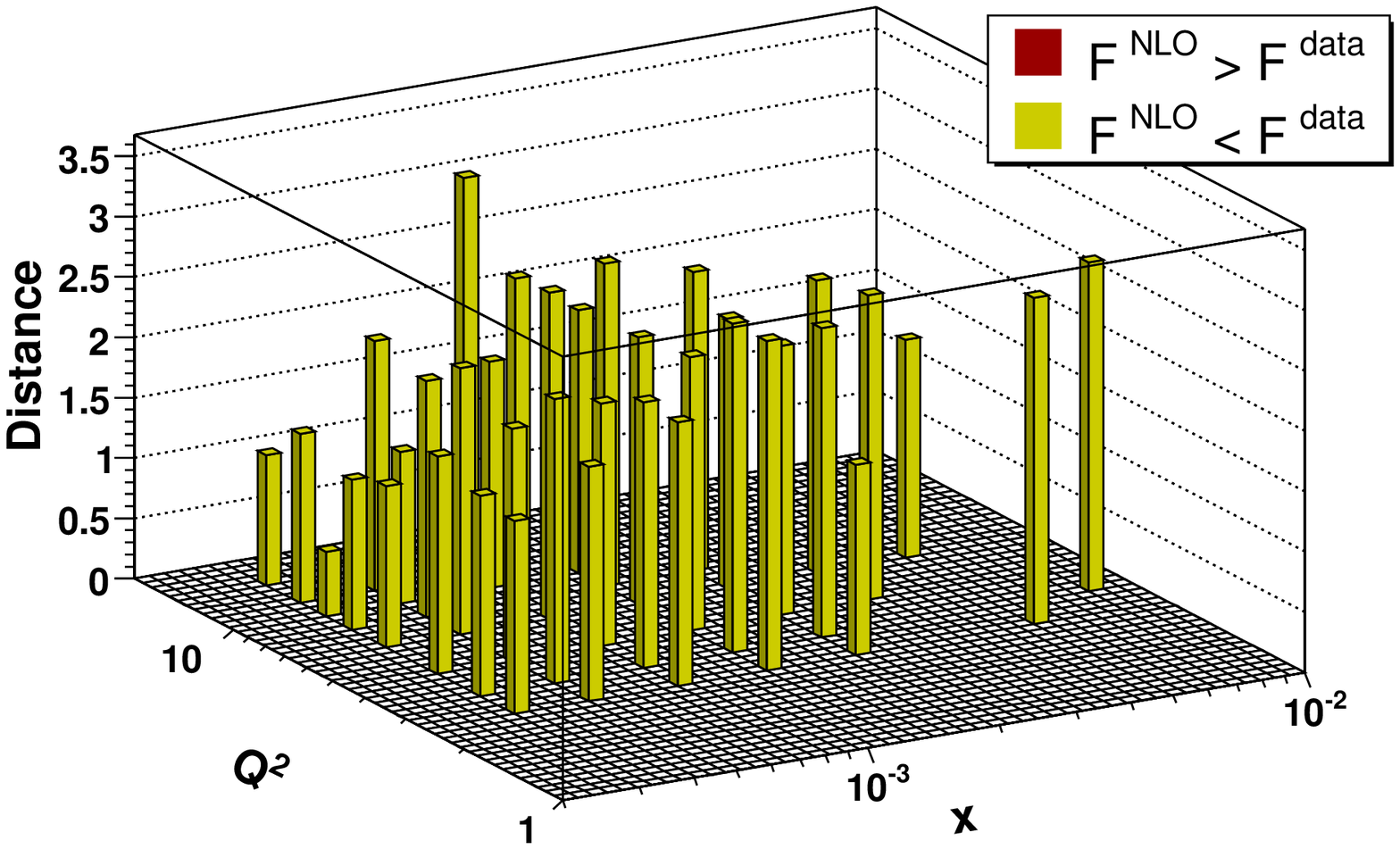, scale=0.32}
\caption{\small 
Statistical distance, Eq.~(\ref{stat_dist}), between small-$x$ HERA data
and NLO DGLAP prediction for  fits without
kinematical cuts (top row) and fits with the 
cut at  $A_{\rm cut}=1.5$ (bottom row). Both results obtained using
NNPDF1.2 with separate HERA data (left) and NNPDF2.0  with combined
HERA data (right) are shown.}
\label{stat_dist_plot}\label{stat_dist_nc_plot}
\end{figure}
In Fig.~\ref{stat_dist_nc_plot} we plot the statistical distance
Eq.~(\ref{stat_dist}) between small-$x$ HERA data
and the NNPDF1.2 and NNPDF2.0 fits without cuts (i.e. with $A_{\rm cut}=0$)
and with the  cut  $A_{\rm cut}=1.5$. Again, we
see that with 
NNPDF1.2 if all data  are included the fit
manages to compensate for the deviation by readjusting the PDFs: the fit
lies both above and below the data and the mean distances is compatible with zero:
 from the points plotted in Fig.~\ref{stat_dist_nc_plot} we obtain
$\left<d_{stat}\right> = 0.06\pm 0.56$.
On the other hand using NNPDF2.0, based on the combined
HERA-I data,
we get $\left<d_{stat}\right> = 1.1 \pm 0.7$, which shows a systematic 
tension at
the one-$\sigma$ level between data and theory. When the cut is
applied, the discrepancy is apparent:
using  NNPDF1.2 
$\left<d_{stat}\right> = 0.95\pm 0.45$, while with NNPDF2.0 
$\left<d_{stat}\right> = 2.0 \pm 0.7$, i.e. a systematic deviation
between data and prediction now at the three-$\sigma$ level.

It is interesting to note that the significance of the effect is
considerably weakened if one instead of performing the cut
Eq.~(\ref{qs}) were to simply cut out the small-$x$ region at all $Q^2$.
For example, if we consider the region $x\le 0.01$
 we obtain (from HERA-I data only) 
$\left<d_{stat}\right> = 1.1 \pm 0.9$ from the 
global fit ($A_{\rm cut}=0$) and $\left<d_{stat}\right> = 1.2 \pm 1.0$ from
the fit with maximum cut ($A_{\rm cut}=1.5$): this shows that indeed
it is only at low $Q^2$ that deviations appear, as one would expect of
an effect driven by perturbative evolution. A recent
study~\cite{Lai:2010vv} did find that in the low-$x$ region the
distance fluctuations are larger than expected, consistent with our
conclusions,
but no significant deviation of $\langle d\rangle$ from zero was
found in this less sensitive $x\le 0.01$ region for an uncut fit, also
consistent with our conclusion.  

\begin{figure}
\centering
\epsfig{file=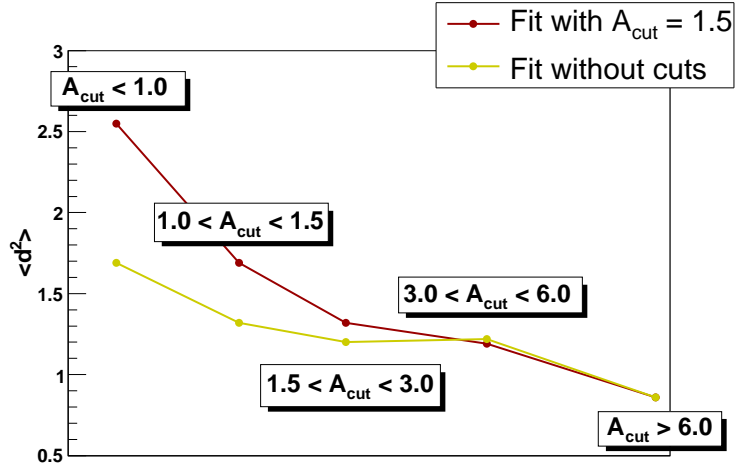, scale=0.5}
\caption{\small $\left< d_{stat}^2 \right>$ computed in the different kinematic 
slices of 
Fig.~\ref{fig:dataplot}: from the NNPDF2.0 fit without
kinematic cuts (yellow, lower curve)  and with the 
$A_{\rm cut}=1.5$ cut
(red, top curve).}\label{plot_d2}
\end{figure}

Evidence for a systematic deviation between data and theory is also
provided by studying the behaviour of 
the $\left< d_{stat}^2\right>$  in different 
kinematic slices, both without cuts and with   $A_{\rm cut}=1.5$.
The results, displayed in Fig.~\ref{plot_d2}, show that 
data and theory increasingly deviate as one moves towards the
small-$x$, small-$Q^2$ region. This deviation is already present when
all data are fitted, but it becomes significantly stronger when the
cut is applied. However, the discrepancy is concentrated in the region
which is affected by the cuts. Indeed,
in Tab.~\ref{table_chi2_cut} we compare the $\chi^2$ of various
datasets for the cut and uncut fits: the quality of the
fit to high--scale hadronic
data, unaffected by the cuts, is the same in the two fits.

\begin{table}
\centering
\begin{tabular}{|c|c|c|}
\hline  
Fit & All dataset &  Only fitted points \\
\hline 
\hline
$\chi^{2}_{\tot}$ & 1.78   & 1.14 \\
\hline
\hline
 NMC-pd    & 0.98 & 0.98\\
\hline
NMC             & 1.75  & 1.75\\
\hline
SLAC            &  1.42 & 1.42\\
\hline
BCDMS           &  1.22 & 1.22\\
\hline
HERAI        &  \emph{4.54} & 1.04\\
\hline
CHORUS          & 1.14 & 1.14\\
\hline
NTVDMN          &  0.70 & 0.70\\
\hline
ZEUS-H2         & 1.23 & 1.23\\
\hline
DYE605          &  0.82 & 0.82 \\
DYE866          &  1.25 & 1.25\\
\hline
CDFWASY         &  1.86 & 1.86\\
\hline
CDFZRAP         &  1.85 & 1.82\\
\hline
D0ZRAP          & 0.56 & 0.56\\
\hline
CDFR2KT         & 0.66 & 0.66\\
\hline
D0R2CON         & 0.82 & 0.82\\
\hline
\end{tabular}
\caption{\small The $\chi^2$ of the
individual experiments included in NNPDF2.0 for the
$A_{\rm cut} = 1.5$ fit. Col. 2 shows $\chi^2$ computed on 
all NNPDF2.0 dataset, while Col. 3 the $\chi^2$ computed
only on data which pass the cut.}
\label{table_chi2_cut}
\end{table}

Having strengthened our previous~\cite{Caola:2009iy} 
conclusion that there is evidence for deviations from NLO
DGLAP evolution in small-$x$ and $Q^2$ HERA data one may ask what are
possible theoretical explanations for the observed effect.
Because NLO DGLAP 
overestimates the amount of evolution required to reproduce
experimental data,
 NNLO corrections as a possible explanation are ruled out, as they
 would lead to yet stronger evolution in this region thus making the
 discrepancy larger. This conclusion was recently confirmed by
the HERAPDF group, which finds that the description
of  small-$x$ and $Q^2$ HERA-I data worsens when NNLO
corrections are included~\cite{mandypdf4lhc}. Charm mass
effects, not included in NNPDF2.0, could be partly responsible, but they
seem~\cite{Caola:2009iy} too small to account for the data. This
conclusion is borne out by preliminary studies based on the
 NNPDF2.1 set~\cite{Rojo:2010gv} which does include charm mass effects
 using the  FONLL~\cite{Forte:2010ta,LH} framework, and which confirm
 the conclusions of the present study.

Interestingly, the small-$x$ resummation corrections shown
in Fig.~\ref{plot_f2resum} and discussed in Sect.~\ref{sec:smallxres}
go in the right direction, and appear to be roughly of the size which
is needed to explain the data. A quantitative confirmation that this
is actually the case could come from a fully resummed
PDF fit, which is doable using current knowledge and it would only
require implementation of the resummation in a PDF fitting code.
An alternative interesting possibility is that the observed slow-down
of perturbative evolution may be due to saturation effects related to
parton recombination. However, it is more difficult to single out a
clear signature for these effects, given that saturation models
usually yield predictions for the $x$ dependence of structure
functions, rather than their scale dependence which is relevant in this
context.

Finally, one may ask whether these deviations, if real, might bias LHC
phenomenology. A first observation is that these deviations might
explain the well--known fact~\cite{Altarelli:2008zz} that the $\alpha_s$
value obtained  from deep--inelastic scattering tends to be lower that
the global average: if the observed evolution is weaker than the
predicted one, the value of the  coupling is biased downwards: the
value of $\alpha_s$ from a fully resummed fit would be higher.
A more direct impact on HERA phenomenology can be assessed by
comparing predictions for LHC standard candles obtained from cut and
uncut fits: their difference provides a conservative upper bound for
the phenomenological impact of these deviations.
In Table~\ref{table_lhc} we show 
results for $W$, $Z$, Higgs 
and $t\bar t$ inclusive
production at the LHC at 7 TeV center of mass energies, computed with the 
MCFM code~\cite{Campbell:2000bg}. Even with
the largest kinematical cuts, $A_{\rm cut}=1.5$, the corrections
are moderate, below 
the 1--sigma level (except for $t\bar{t}$), of similar size of other
comparable effects at the precision level, such as
$\alpha_s$ uncertainties~\cite{Demartin:2010er}
or variations of the charm mass.  

The impact of the effect we discovered is 
moderate at present but it might become significant as the accuracy of PDF
determination improves. It will be interesting to see whether further
confirmation of the effect comes from other groups. Its full understanding
might lead to a deeper grasp of perturbative QCD.

\begin{table}
\begin{center}
\begin{tabular}{|c|c|c|}
\hline
Observable & NNPDF2.0 without cuts & NNPDF2.0 with $A_{\rm cut} = 1.5$\\
\hline
\hline
$\sigma(W^+) \text{B}_{l^+\nu}$ [nb] & 5.80 $\pm$ 0.09 & 5.87 $\pm$ 0.13 \\
$\sigma(W^-) \text{B}_{l^-\bar\nu}$ [nb] & 3.97 $\pm$ 0.06 & 4.01 $\pm$ 0.07 \\
$\sigma(Z) \text{B}_{l^+ l^-}$ [nb] & 2.97 $\pm$ 0.04 & 2.98 $\pm$ 0.05 \\
$\sigma(t\bar t)$ [pb] & 169 $\pm$ 5 & 160 $\pm$ 7 \\
$\sigma(H, m_H = 120~\text{GeV})$ [pb] & 11.60 $\pm$ 0.15 & 11.53 $\pm$ 0.25\\
\hline
\end{tabular}
\end{center}
\caption{\small LHC observables at 7 TeV computed 
from the default NNPDF2.0 set and with the fit
with kinematical cut $A_{\rm cut}=1.5$ using
MCFM~\cite{Campbell:2000bg}. 
All PDF uncertainties are
1--sigma. }
\label{table_lhc}
\end{table}

\bigskip
\bigskip
\begin{center}
\rule{3cm}{.1pt}
\end{center}
\bigskip
\bigskip
The NNPDF2.0 PDFs (sets of $N_{\rm rep}=100$ and 1000 replicas) 
and the PDF sets based on NNPDF2.0 with various
$A_{\rm cut}$ kinematical cuts are 
available at the
NNPDF web site,
\begin{center}
{\bf \url{http://sophia.ecm.ub.es/nnpdf}~.}
\end{center}
NNPDF2.0 is  also available
through the LHAPDF interface~\cite{Bourilkov:2006cj}.






\bibliographystyle{h-elsevier3}
\bibliography{biblio}

\begin{thebibliography}{10}

\bibitem{Altarelli:2008zz}
G. Altarelli,
\newblock In *Landolt-Boernstein I 21A: Elementary particles* 4.

\bibitem{Catani:1990eg}
S. Catani, M. Ciafaloni and F. Hautmann,
\newblock Nucl. Phys. B366 (1991) 135.

\bibitem{Catani:1994sq}
S. Catani and F. Hautmann,
\newblock Nucl. Phys. B427 (1994) 475, hep-ph/9405388.

\bibitem{Altarelli:2003hk}
G. Altarelli, R.D. Ball and S. Forte,
\newblock Nucl. Phys. B674 (2003) 459, hep-ph/0306156.

\bibitem{Altarelli:2005ni}
G. Altarelli, R.D. Ball and S. Forte,
\newblock Nucl. Phys. B742 (2006) 1, hep-ph/0512237.

\bibitem{Altarelli:2008aj}
G. Altarelli, R.D. Ball and S. Forte,
\newblock Nucl. Phys. B799 (2008) 199, 0802.0032.

\bibitem{Ciafaloni:2003rd}
M. Ciafaloni et~al.,
\newblock Phys. Rev. D68 (2003) 114003, hep-ph/0307188.

\bibitem{Ciafaloni:2003kd}
M. Ciafaloni et~al.,
\newblock Phys. Lett. B587 (2004) 87, hep-ph/0311325.

\bibitem{Ciafaloni:2007gf}
M. Ciafaloni et~al.,
\newblock JHEP 08 (2007) 046, 0707.1453.

\bibitem{Ball:2001pq}
R.D. Ball and R.K. Ellis,
\newblock JHEP 05 (2001) 053, hep-ph/0101199.

\bibitem{Marzani:2008az}
S. Marzani et~al.,
\newblock Nucl. Phys. B800 (2008) 127, 0801.2544.

\bibitem{Marzani:2008ih}
S. Marzani et~al.,
\newblock Nucl. Phys. Proc. Suppl. 186 (2009) 98, 0809.4934.

\bibitem{Marzani:2008uh}
S. Marzani and R.D. Ball,
\newblock Nucl. Phys. B814 (2009) 246, 0812.3602.

\bibitem{Marzani:2009hu}
S. Marzani and R.D. Ball,
\newblock (2009), 0906.4729.

\bibitem{Diana:2009xv}
G. Diana,
\newblock Nucl. Phys. B824 (2010) 154, 0906.4159.

\bibitem{Diana:2010ef}
G. Diana, J. Rojo and R.D. Ball,
\newblock (2010), 1006.4250.

\bibitem{Ball:2007ra}
R.D. Ball,
\newblock Nucl. Phys. B796 (2008) 137, 0708.1277.

\bibitem{Rojo:2010gv}
J. Rojo et~al.,
\newblock (2010), 1007.0354.

\bibitem{Ball:2010de}
R.D. Ball et~al.,
\newblock Nucl. Phys. B838 (2010) 136, 1002.4407.

\bibitem{Nadolsky:2008zw}
P.M. Nadolsky et~al.,
\newblock Phys. Rev. D78 (2008) 013004, 0802.0007.

\bibitem{Lai:2010nw}
H.L. Lai et~al.,
\newblock (2010), 1004.4624.

\bibitem{Martin:2009iq}
A.D. Martin et~al.,
\newblock Eur. Phys. J. C63 (2009) 189, 0901.0002.

\bibitem{f2ns}
S. Forte et~al.,
\newblock JHEP 05 (2002) 062, hep-ph/0204232.

\bibitem{f2p}
The NNPDF Collaboration, L. Del~Debbio et~al.,
\newblock JHEP 03 (2005) 080, hep-ph/0501067.

\bibitem{DelDebbio:2007ee}
The NNPDF Collaboration, L. Del~Debbio et~al.,
\newblock JHEP 03 (2007) 039, hep-ph/0701127.

\bibitem{Ball:2008by}
The NNPDF Collaboration, R.D. Ball et~al.,
\newblock Nucl. Phys. B809 (2009) 1, 0808.1231.

\bibitem{Rojo:2008ke}
The NNPDF Collaboration, J. Rojo et~al.,
\newblock (2008), 0811.2288.

\bibitem{Ball:2009mk}
The NNPDF Collaboration, R.D. Ball et~al.,
\newblock Nucl. Phys. B823 (2009) 195, 0906.1958.

\bibitem{Ball:2009qv}
The NNPDF Collaboration, R.D. Ball et~al.,
\newblock JHEP 05 (2010) 075, 0912.2276.

\bibitem{Dittmar:2009ii}
M. Dittmar et~al.,
\newblock (2009), 0901.2504.

\bibitem{Stasto:2000er}
A.M. Stasto, K.J. Golec-Biernat and J. Kwiecinski,
\newblock Phys. Rev. Lett. 86 (2001) 596, hep-ph/0007192.

\bibitem{Caola:2008xr}
F. Caola and S. Forte,
\newblock Phys. Rev. Lett. 101 (2008) 022001, 0802.1878.

\bibitem{H1ZEUS:2009wt}
H1, F.D. Aaron et~al.,
\newblock JHEP 01 (2010) 109, 0911.0884.

\bibitem{Ball:1994du}
R.D. Ball and S. Forte,
\newblock Phys. Lett. B335 (1994) 77, hep-ph/9405320.

\bibitem{Caola:2009iy}
F. Caola, S. Forte and J. Rojo,
\newblock Phys. Lett. B686 (2010) 127, 0910.3143.

\bibitem{Lai:2010vv}
H.L. Lai et~al.,
\newblock (2010), 1007.2241.

\bibitem{mandypdf4lhc}
A. Cooper-Sarkar,
\newblock {http://indico.cern.ch/conferenceDisplay.py?confId=98883},
\newblock 2010.

\bibitem{Forte:2010ta}
S. Forte et~al.,
\newblock Nucl. Phys. B834 (2010) 116, 1001.2312.

\bibitem{LH}
SM and NLO Multileg Working Group, J.R. Andersen et~al.,
\newblock (2010), 1003.1241.

\bibitem{Campbell:2000bg}
J.M. Campbell and R.K. Ellis,
\newblock Phys. Rev. D62 (2000) 114012, hep-ph/0006304.

\bibitem{Demartin:2010er}
F. Demartin et~al.,
\newblock Phys. Rev. D82 (2010) 014002, 1004.0962.

\bibitem{Bourilkov:2006cj}
D. Bourilkov, R.C. Group and M.R. Whalley,
\newblock (2006), hep-ph/0605240.

\end{thebibliography}







\end{document}